\documentclass[manuscript,natbib=false,screen]{acmart}

\AtBeginDocument{%
  }

\setcopyright{acmlicensed}
\copyrightyear{2026}
\acmYear{2026}
\acmDOI{XXXXXXX.XXXXXXX}

\acmJournal{JACM}
\acmVolume{37}
\acmNumber{4}
\acmArticle{111}
\acmMonth{8}

\RequirePackage[
  datamodel=acmdatamodel,
  style=acmnumeric,
  ]{biblatex}

\usepackage{booktabs}
\usepackage{graphicx}
\usepackage{multirow}
\usepackage{array}
\usepackage{xcolor}
\usepackage{adjustbox}
\usepackage{enumitem}
\usepackage{makecell}

\begin{document}

% \title[The Evolving Role of Large Language Models in Scientific Innovation]{The Evolving Role of Large Language Models in Scientific Innovation: Assistant, Collaborator, Scientist, and Evaluator}

\title[Evolving Roles of LLMs in Scientific Innovation]{Evolving Roles of LLMs in Scientific Innovation: Assistant, Collaborator, Scientist, and Evaluator}

\author{Haoxuan Zhang}
\orcid{0000-0003-2654-6127}
\email{haoxuanzhang@my.unt.edu}
\affiliation{%
  \institution{Department of Information Science, University of North Texas}
  \city{Denton}
  \state{Texas}
  \country{USA}
}

\author{Ruochi Li}
\orcid{0009-0006-6677-5222}
\email{rli14@ncsu.edu}
\affiliation{%
  \institution{Department of Computer Science, North Carolina State University}
  \city{Raleigh}
  \state{North Carolina}
  \country{USA}
}

\author{Yang Zhang}
\orcid{0000-0001-6821-2710}
\email{yang.zhang@unt.edu}
\affiliation{%
  \institution{Department of Data Science, University of North Texas}
  \city{Denton}
  \state{Texas}
  \country{USA}
}

\author{Ting Xiao}
\orcid{0000-0001-8548-5710}
\email{ting.xiao@unt.edu}
\affiliation{%
  \institution{Department of Data Science, University of North Texas}
  \city{Denton}
  \state{Texas}
  \country{USA}
}

\author{Jiangping Chen}
\orcid{0000-0002-7016-4583}
\email{jpchen@illinois.edu}
\affiliation{%
  \institution{School of Information Sciences, University of Illinois Urbana-Champaign}
  \city{Champaign}
  \state{Illinois}
  \country{USA}
}

\author{Junhua Ding}
\orcid{0000-0002-2129-1586}
\email{junhua.ding@unt.edu}
\affiliation{%
  \institution{Department of Data Science, University of North Texas}
  \city{Denton}
  \state{Texas}
  \country{USA}
}

\author{Haihua Chen}
\orcid{0000-0002-7088-9752}
\authornote{Corresponding author.}
\email{haihua.chen@unt.edu}
\affiliation{%
  \institution{Department of Data Science, University of North Texas}
  \city{Denton}
  \state{Texas}
  \country{USA}
}
\renewcommand{\shortauthors}{Zhang et al.}

\begin{abstract}
Large language models (LLMs) are increasingly used in scientific research and discovery, supporting tasks ranging from literature retrieval and synthesis to hypothesis generation, autonomous experimentation, and research evaluation. Existing surveys often conflate scientific research with scientific discovery and typically organize systems by domain, task, or autonomy level alone. In this survey, we propose a four-role framework for understanding LLMs in scientific innovation: \textbf{Assistant}, \textbf{Collaborator}, \textbf{Scientist}, and \textbf{Evaluator}. The framework integrates three complementary dimensions—autonomy level, cognitive function, and scientific innovation—to distinguish research-oriented support from frontier-oriented discovery. We review representative methods, benchmarks, and evaluation practices for each role, examining their capabilities, limitations, and human oversight requirements. Across the literature, \textbf{Assistant systems} are comparatively mature in retrieval and synthesis but remain unreliable in open-ended applications; \textbf{Collaborator systems} expand the space of candidate hypotheses yet struggle with novelty-grounding trade-offs; \textbf{Scientist systems} increasingly automate research workflows but face reliability and safety bottlenecks; and \textbf{Evaluator systems} support review and verification while remaining weak in novelty assessment. We argue that progress in AI for science depends not only on model capability, but also on evaluation, oversight, accountability, and institutional integration.

\end{abstract}






\maketitle

\section{Introduction}\label{sec1}

Recent progress in artificial intelligence (AI), especially large language models (LLMs) has intensified interest in their role in scientific innovation. In this survey,  \textbf{scientific innovation} refers to two related but analytically distinct activities: \textbf{scientific research}, which evaluates and refines claims through systematic investigation, and \textbf{scientific discovery}, which introduces new hypotheses, mechanisms, or phenomena that extend current knowledge~\cite{wang2023scientific,reddy2025towards,chen2025ai4research}. This distinction is useful because they involve different objectives, workflows, and evaluation criteria. At the same time, contemporary science faces a well-documented tension between expanding publication volume and the more limited growth of science's cognitive extent, alongside incentives that favor conservative strategies over higher-risk exploration~\cite{milojevic2015quantifying,foster2015tradition}. These conditions increase the burden of synthesis and may inhibit the cross-disciplinary connections often associated with breakthrough innovation~\cite{wang2024exploring}.

\textbf{Distinction Between Existing Surveys.} LLMs are increasingly being used across the scientific innovation, supported by advances in alignment, reasoning, tool use, multimodal modeling, and multi-agent coordination~\cite{bai2022training,wei2022chain,yao2022react,xu2023multimodal,huang2024detection,tran2025multi,besta2025reasoning}. Existing surveys have approached this landscape from complementary angles: stage-based surveys structure coverage around sequential pipeline steps (hypothesis formulation, experiment planning, writing, and review), treating all stages as structurally equivalent without differentiating the cognitive demands or oversight requirements that separate different forms of scientific participation~\cite{gao2024empowering,luo2025llm4sr,zhou-etal-2025-hypothesis,zhang2025advancing,alkan2025survey}; autonomy-based surveys arrange systems along a single dimension from tool use to agentic behavior, but autonomy alone does not determine how a system should be evaluated or governed, and validation tasks are typically excluded from scope~\cite{zheng2025automation,wei2025agentic}; and~\textcite{chen2025ai4research} introduce a five-task taxonomy that spans the research lifecycle and includes a discovery category, but treat research and discovery as separable pipeline stages rather than as analytically distinct activities imposing different evaluation criteria across all tasks. Domain-based and comprehensive surveys provide empirical breadth without a principled taxonomy~\cite{zhang2024comprehensive,yu2025empowering,reddy2025towards}. Across these approaches, the systematic variation of oversight requirements with system role remains underexamined, and the boundary between research support and frontier discovery is rarely operationalized as an evaluative criterion.

Therefore, motivated by these limitations, this survey identifies three specific gaps in the current literature.

\textbf{Issue 1: Conflation of Research and Discovery.} Scientific research and scientific discovery require different methods and should be evaluated against different criteria~\cite{wang2023scientific,reddy2025towards}. However, existing surveys often discuss them under a common capability rubric. This weakens interpretation: autonomous research benchmarks typically emphasize process correctness and workflow completion, whereas discovery-oriented systems must also be evaluated in terms of novelty, external validity, and contribution to the frontier of knowledge. Without this distinction, routine workflow automation and genuine discovery can be placed on the same scale in ways that obscure present capability boundaries.

\textbf{Issue 2: Lack of a Theoretically Grounded Role Taxonomy.} Existing surveys usually organize LLM systems by domain, task type, or degree of autonomy~\cite{luo2025llm4sr,chen2025ai4research,zheng2025automation}. These categorizations are useful descriptively, but they provide weaker guidance on why role boundaries matter, which cognitive demands separate different forms of scientific participation, and how oversight requirements change as autonomy increases. A more principled taxonomy can be developed by integrating perspectives from human-centered AI~\cite{shneiderman2020human}, cognitive complexity~\cite{krathwohl2002revision}, and scientific paradigm theory~\cite{kuhn1970structure}.

\textbf{Issue 3: Underexamined Human Oversight Across Roles.} Human oversight or supervision is often treated as a residual implementation detail rather than as a role-specific design requirement. Yet the literature points to different failure modes at different capability levels: hallucinated references and unsupported synthesis in assistant-style use~\cite{messeri2024artificial,luo2025llm4sr}; trade-offs among novelty, feasibility, and factual grounding in collaborator-style hypothesis generation~\cite{liu2025hypobench,xiong2025truthhypo}; compounding failures and unauthorized actions in autonomous research settings~\cite{kon2026expbench,boiko2023autonomous,mandal2025evaluating}; and over-leniency or weak novelty calibration in evaluator-style review systems~\cite{zhou2024llm,li2025unveiling,russo2025ai}. These patterns suggest that oversight should be analyzed as an integral dimension of \textbf{scientific AI systems} (LLM-based systems configured to support or execute scientific tasks, from retrieval and synthesis to autonomous experimentation and evaluation~\cite{chen2025ai4research,luo2025llm4sr,lu2026ai}) rather than as an afterthought.

To address these issues, we propose a four-role framework for LLMs in scientific innovation that organizes existing systems into Assistant, Collaborator, Scientist, and Evaluator. By making role differences explicit, this framework provides the conceptual basis for comparing current systems, analyzing their limitations, and identifying priorities for future research. Our contributions are as follows:

\begin{enumerate}[leftmargin=*]
\item We maintain a systematic distinction between scientific research and scientific discovery throughout the survey, and use this distinction to clarify role-specific evaluation criteria.

\item We propose a four-role framework (\textbf{Assistant}, \textbf{Collaborator}, \textbf{Scientist}, and \textbf{Evaluator}) grounded in three complementary dimensions: autonomy level, cognitive function, and scientific innovation. This framework supports more precise comparison of LLM systems across different forms of scientific participation.

\item We synthesize role-specific failure modes, oversight requirements, and ethical considerations, with particular attention to the stages at which human judgment remains necessary for reliable and accountable scientific practice.
\end{enumerate}

The remainder of this survey is organized as follows. Section~\ref{sec:roledefine} introduces the four-role taxonomy and its theoretical foundations. Sections~\ref{Assist}–\ref{llmasval} examine the four roles in turn: Assistant (scientific research), Collaborator (scientific discovery), Scientist (autonomous research and discovery), and Evaluator (scientific validation).  Section~\ref{discussion} synthesizes cross-role insights and discusses implications for human agency, system reliability, and scientific integrity, and outlines directions for future research.

\section{The Role Definition of LLMs in Scientific Innovation}
\label{sec:roledefine}

\begin{figure*}[t]
\centering
\includegraphics[width=0.9\textwidth]{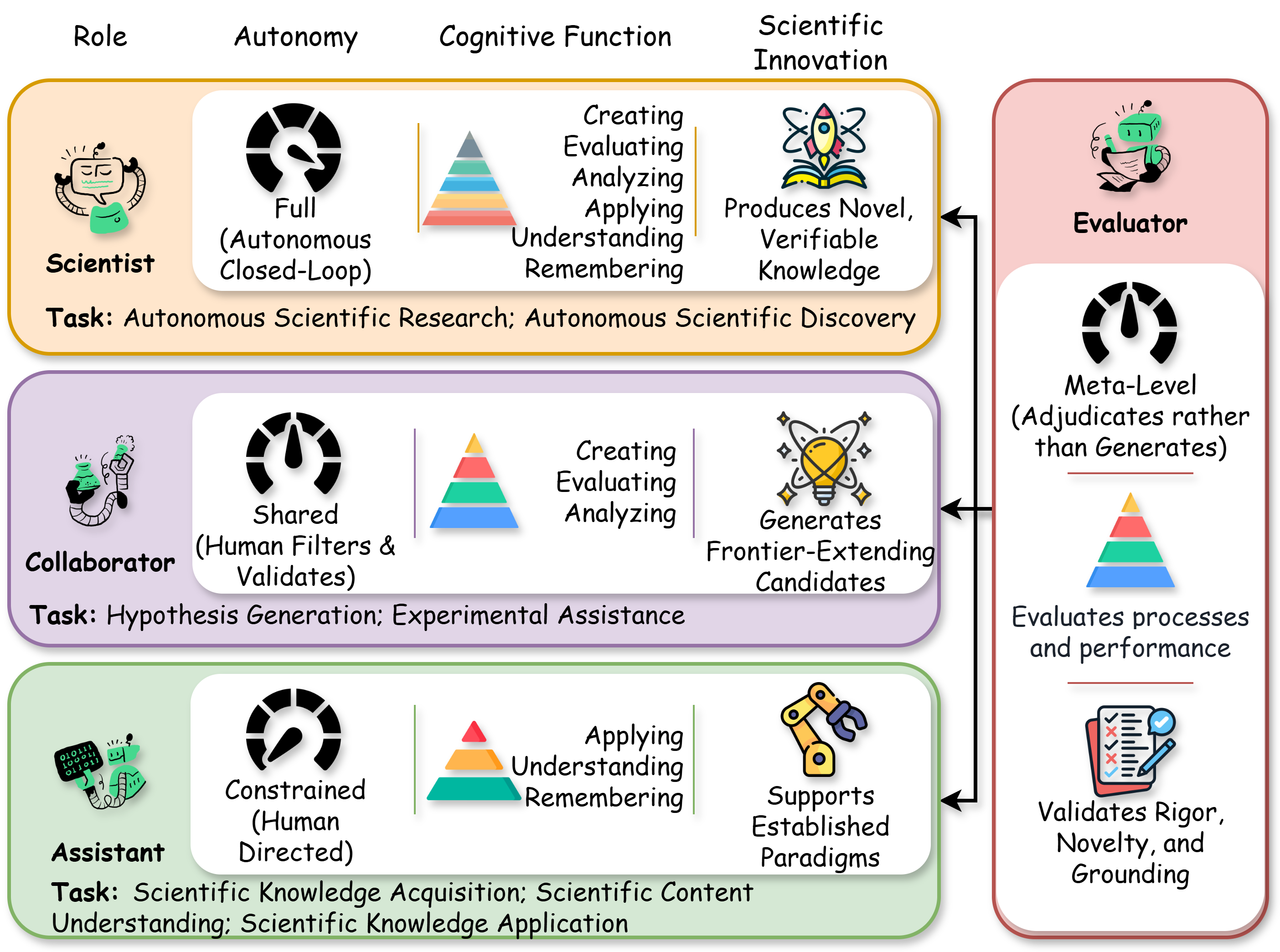}
\caption{The four-role framework for LLMs in scientific innovation, organized along three dimensions: autonomy level, cognitive function, and scientific innovation. Assistant, Collaborator, and Scientist form a progression in generative scientific participation, from constrained workflow support to shared-autonomy hypothesis generation to higher-autonomy closed-loop discovery. Evaluator occupies an orthogonal meta-level role, assessing the rigor, novelty, significance, and grounding of outputs produced by the other three roles.}
\label{fig:four_role_framework}
\end{figure*}

In this survey, we use \textbf{scientific innovation} as an umbrella term covering two related but distinct activities: \textit{scientific research}, which concerns how effectively a system supports or executes the research process, and \textit{scientific discovery}, which concerns whether a system contributes knowledge that is genuinely new and externally verifiable. This distinction matters because a system may perform well on research-oriented tasks such as retrieval, synthesis, planning, or execution while remaining much less capable on discovery-oriented tasks that require novelty, robustness, and acceptance under external scrutiny.

Existing surveys typically organize LLM systems by domain, application type, or degree of autonomy~\cite{luo2025llm4sr,chen2025ai4research,zheng2025automation}. These categorizations are useful descriptively, but they provide less guidance on what kind of scientific participant a system is, what cognitive work it performs, and where human oversight remains necessary. To address this gap, we define roles along three complementary dimensions: \textbf{autonomy level}, which specifies how independently a system acts within a scientific workflow~\cite{shneiderman2020human}; \textbf{cognitive function}, which specifies the level of reasoning required of the system, drawing on the six-level hierarchy of Bloom's revised taxonomy (Remember, Understand, Apply, Analyze, Evaluate, Create)~\cite{krathwohl2002revision}; and \textbf{scientific innovation}, which specifies whether the system primarily supports established inquiry or contributes more directly to frontier-oriented discovery~\cite{kuhn1970structure}. Figure~\ref{fig:four_role_framework} summarizes how the three dimensions introduced above jointly structure the survey. Assistant, Collaborator, and Scientist form a progression in generative scientific participation, corresponding to increasing autonomy, increasing cognitive demand, and increasing potential contribution to discovery, whereas Evaluator occupies an orthogonal meta-level role devoted to assessing the rigor, novelty, significance, and grounding of outputs produced by the other roles. Table~\ref{tab:role_taxonomy} complements this visual overview by listing the defining properties and representative tasks of each role in compact form.

\begin{table*}[htbp]
\centering
\scriptsize
\setlength{\extrarowheight}{4pt}
\caption{The Four-Role Taxonomy for LLMs in Scientific Innovation.}
\label{tab:role_taxonomy}
\begin{adjustbox}{max width=\textwidth}
\begin{tabular}{@{}p{1.8cm}p{2.0cm}p{2.4cm}p{2.2cm}p{2.2cm}p{3.6cm}@{}}
\toprule
\multirow{2}{*}{\textbf{Role}} &
\multirow{2}{*}{\textbf{Autonomy}} &
\multirow{2}{*}{\textbf{Cognitive Function}} &
\multicolumn{2}{c}{\textbf{Scientific Innovation}} &
\multirow{2}{*}{\textbf{Representative Tasks}} \\
\cmidrule(lr){4-5}
& & & \textbf{Research} & \textbf{Discovery} & \\
\midrule

\textbf{Assistant}
& Constrained; human directs all goals and verifies all outputs
& Remembering, Understanding, Applying
& Supports research execution within established paradigms
& Indirect; does not advance the frontier
& Knowledge extraction and retrieval, scientific QA, survey and taxonomy generation \\

\addlinespace

\textbf{Collaborator}
& Shared; human filters, grounds, and validates
& Analyzing, Evaluating, Creating
& Indirect; does not execute full research workflows
& Candidate-level; generates frontier-extending hypotheses and empirical artifacts pending human verification
& Hypothesis generation, experiment design and optimization, automated lab execution \\

\addlinespace

\textbf{Scientist}
& Full; autonomous closed-loop with minimal human co-direction
& All six levels in autonomous Plan--Act--Observe--Refine cycles
& Pipeline-complete; autonomous workflow from problem formulation to manuscript
& Frontier-advancing; produces externally verifiable novel knowledge
& End-to-end autonomous research, self-driving laboratories, evolutionary program synthesis \\

\addlinespace

\textbf{Evaluator}
& Meta-level; adjudicates rather than generates
& Evaluating
& Assesses manuscript quality, methodological rigor, and peer review validity
& Verifies validity, grounding, and novelty of empirical claims and findings
& Automated peer review, meta-review and rebuttal synthesis, claim verification, novelty and impact assessment \\

\bottomrule
\end{tabular}
\end{adjustbox}
\end{table*}

The taxonomy is intended as an analytical framework for system-task configurations rather than as a fixed labeling scheme for base models. The same underlying model may function as an Assistant in one setting, as a Collaborator in another, or as part of a Scientist-style pipeline when deployed with greater autonomy and tool access~\cite{zheng2025automation,chen2025ai4research,luo2025llm4sr}. Its purpose is therefore to clarify how different combinations of autonomy, cognitive demand, scientific innovation, and oversight requirement correspond to different forms of scientific participation.

\section{LLMs as Assistants: Enabling Scientific Research}
\label{Assist}

As defined in Section~\ref{sec:roledefine}, the Assistant role refers to settings in which LLMs support scientific research under close human direction and verification. In this role, models are primarily used for retrieval, extraction, question answering, and synthesis tasks that operate within established research workflows rather than extending the knowledge frontier.

\subsection{Role Definition}
\label{llm-assit-task}

\begin{figure*}[t]
    \centering
    \includegraphics[width=0.9\linewidth]{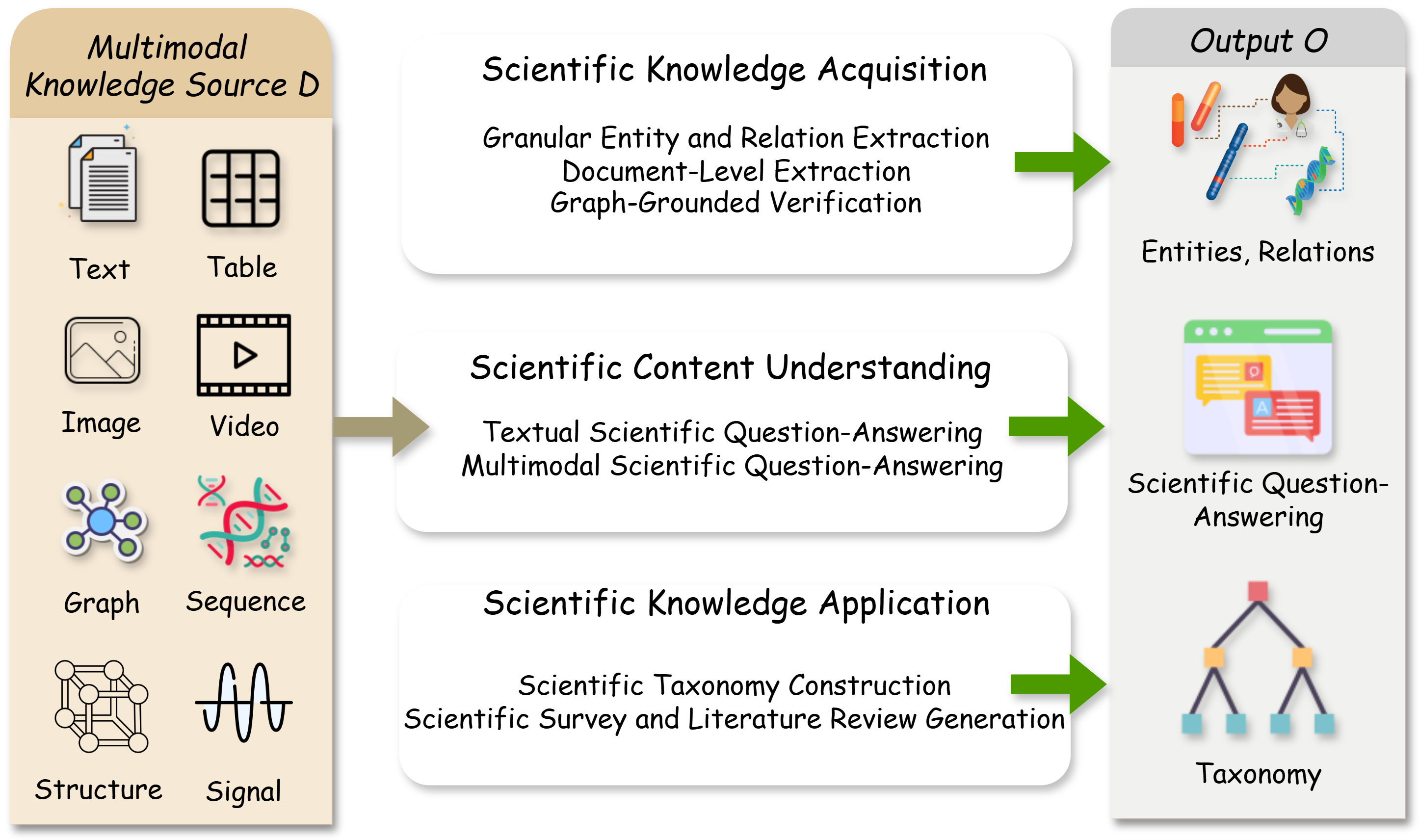}
    \caption{
        The framework of LLMs as assistants in scientific research. Given a scientific query and multimodal knowledge sources, the assistant supports scientific knowledge acquisition, scientific content understanding, and scientific knowledge application, producing outputs such as entities and relations, question-answering responses, taxonomies, and literature surveys.
    }
    \label{fig:assistant-framework}
\end{figure*}

We formalize the LLM-as-Assistant role through the following definition:
\begin{equation}
    \text{LLM}_{\text{Assistant}}(Q,\, \mathcal{D},\, \theta) \;\rightarrow\; O \;\in\; \{O_{\mathrm{ret}},\, O_{\mathrm{ans}},\, O_{\mathrm{art}}\}
    \label{eq:assistant}
\end{equation}
where $Q$ is the scientific query, $\mathcal{D}$ the multimodal knowledge source, $\theta$ the model parameters, and $O$ the resulting output: a retrieved entity or passage ($O_{\mathrm{ret}}$), a reasoned answer ($O_{\mathrm{ans}}$), or a synthesized artifact such as a taxonomy or survey ($O_{\mathrm{art}}$), determined by the cognitive demand of $Q$. Here $\mathcal{D}$ encompasses eight modality types: \textit{Text}, \textit{Table}, \textit{Image}, \textit{Video}, \textit{Graph}, \textit{Sequence} (e.g., DNA, SMILES), \textit{Structure} (3D spatial configurations), and \textit{Signal} (continuous waveforms and spectra), collectively covering the full range of data formats encountered in scientific domains. This formulation unifies extraction, comprehension, and application under a single knowledge-processing view. Figure~\ref{fig:assistant-framework} illustrates the three-layer structure of this framework, mapping each task category to its corresponding output type.

\subsection{Tasks and Approaches}

\subsubsection{Scientific Knowledge Acquisition}
\label{sec:remembering}

At the \textit{Remembering} level, LLM Assistants support scientific knowledge acquisition by locating, extracting, and structuring targeted information from scientific sources. The output $O$ is a retrieved fact, named entity, relational triple, or structured passage extracted directly from $\mathcal{D}$.

\paragraph{(1) Granular Entity and Relation Extraction}
Scientific knowledge acquisition begins with identifying core concepts and relations in scientific text. Earlier sentence-level approaches struggle with long-range dependencies, motivating document-level and jointly modeled extraction frameworks~\cite{zhang2024scier,duan2025scinlp}. In high-volume domains such as biomedicine, systems such as PubTator 3.0 show how large-scale neural pipelines can sustain continuous extraction over millions of documents~\cite{wei2024pubtator}. A related line of work moves beyond span extraction toward transforming scientific prose directly into structured knowledge artifacts such as tables and database-ready records~\cite{dagdelen2024structured,wang2025scidasynth}.

\paragraph{(2) Document-Level Extraction}
Document-level approaches shift the target from local entities to larger scientific structures, such as contribution types, workflow dimensions, argumentation patterns, and future-work statements~\cite{zhang2025massw,chao2023joint,zhang2023automatic}. To support such tasks on full papers, current methods either train explicitly for long-context scientific inputs or distribute extraction across multiple specialized agents before consolidating outputs~\cite{hilgert2024evaluating,song2025scientific}.

\paragraph{(3) Graph-Grounded Verification}
As extraction becomes broader and more open-ended, reliability increasingly depends on external grounding. Recent systems therefore integrate hierarchical retrieval structures, graph-enhanced reasoning, knowledge-graph revision loops, and citation-network-based retrieval to verify extracted content and improve factual consistency~\cite{chen2024hiqa,jia2025hetgcot,xu2025chatpd,sukgarevion,garikaparthi2025mir}. Across these approaches, the key trend is the same: structured knowledge sources are no longer passive repositories, but active constraints on extraction and reasoning.

\subsubsection{Scientific Content Understanding}

At the \textit{Understanding} level, the cognitive demand shifts from locating information to \textit{interpreting} it: given a specific question $Q$, the model must reason over $\mathcal{D}$ to derive a direct answer as output $O$, rather than retrieve a verbatim passage. Approaches are organized into two categories: textual scientific QA, where answers are derived through multi-step reasoning over scientific prose, and multimodal scientific QA, where answers require decoding structured visual artifacts and fusing heterogeneous modalities. 

\paragraph{(1) Textual Scientific QA}
Scientific QA requires models to integrate evidence across passages and documents rather than retrieve isolated spans. Progress has followed a common trajectory: domain adaptation improves technical reading competence~\cite{li2025scilitllm}; retrieval-aware models learn when and how to search external evidence~\cite{jeong2024improving}; and more recent systems target multi-document synthesis, attribution, contradiction handling, and agentic literature search at scale~\cite{skarlinski2024language,asai2026synthesizing,singh2025ai2,he2025pasa}. The main challenge is no longer passage retrieval alone, but evidence integration under attribution and grounding constraints.

\paragraph{(2) Multimodal Scientific QA}
Multimodal scientific QA extends this challenge to charts, tables, figures, and full documents. Research in this area shows that explicit reasoning improves multimodal answer quality, whether through thought chains, staged cross-modal inference, or external knowledge grounding~\cite{lu2022learn,zhang2023multimodal,mondal2024kam}. More specialized systems target chart reasoning, table understanding, and whole-document multimodal QA through instruction tuning, structural reasoning, and unified multimodal encoders~\cite{liu2023matcha,masry2024chartinstruct,masry2025chartgemma,he2025distill,wang2024chain,zheng2024multimodal,yin2025enhancing,hu2024mplug,li2024m3sciqa,singh2024scidqa}. The main bottleneck remains reliable cross-modal grounding rather than raw linguistic fluency.

\subsubsection{Scientific Knowledge Application}

At the \textit{Applying} level, LLM Assistants move beyond interpreting existing content to \textit{constructing structured knowledge artifacts} through cross-source synthesis. The output $O$ is an artifact, such as a hierarchical taxonomy or literature survey, that reorganizes and consolidates existing knowledge into a coherent whole. The open-ended nature of this task distinguishes it from Understanding: rather than answering what $\mathcal{D}$ says, the model synthesizes across $\mathcal{D}$ to produce an artifact that did not previously exist in that organized form, without introducing claims that extend beyond the source material.

\paragraph{(1) Scientific Taxonomy Construction}
Taxonomy construction requires LLMs to organize a field into coherent conceptual hierarchies rather than interpret individual papers in isolation. Existing work explores graph-based hierarchy induction, LLM-guided multi-aspect clustering, and structured table synthesis to make scientific literatures more navigable~\cite{kang2024researcharena,hsu2024chime,zhu2025context,newman2024arxivdigestables,wang2025can}. The central challenge is balancing interpretability with sufficient granularity.

\paragraph{(2) Scientific Survey and Literature Review Generation}
Automated survey generation combines coverage, synthesis, and factual grounding. Systems in this space differ mainly in how they manage retrieval, outlining, multi-stage drafting, and quality control~\cite{agarwal2024litllm,wang2024autosurvey,liang2025surveyx,yan2025surveyforge,asai2026synthesizing,bao2025surveygen,wu2025automated,li2025chatcite,zhang2025mixture}. Across these approaches, the main unresolved problem is not producing fluent long-form text, but maintaining structural coherence, citation fidelity, and analytical depth as synthesis scales.

\subsection{Benchmarks and Evaluation}

Table~\ref{tab:ass_benchmarks} presents representative benchmarks organized across the three cognitive levels established in Section~\ref{llm-assit-task}, reflecting a progressive shift from extraction-precision metrics at the \textit{Acquisition} level toward semantic scoring at the \textit{Understanding} level and multi-dimensional human judgment at the \textit{Application} level.

\begin{table*}[htbp]
\centering
\scriptsize
\setlength{\extrarowheight}{4pt}
\caption{Representative Benchmarks for LLMs as Assistants.}
\label{tab:ass_benchmarks}
\begin{adjustbox}{max width=\textwidth}
\begin{tabular}{@{}m{0.5cm}p{2.1cm}cp{3.2cm}p{4.7cm}p{3cm}@{}}
\toprule
\textbf{Level} & \textbf{Benchmark} & \textbf{Year} & \textbf{Scale} & \textbf{Domain} & \textbf{Eval.\ Metrics} \\
\midrule
\multirow{4}{*}{\centering\rotatebox[origin=c]{90}{\textbf{Acquisition}}}
& PubTator 3.0~\cite{wei2024pubtator} & 2024 & 36M abstracts / 6M full-text articles & Biomedical NER (6 types) and RE (12 types) over PubMed & Entity F\textsubscript{1}, Relation F\textsubscript{1} \\
& SciNLP~\cite{duan2025scinlp} & 2025 & 60 papers / 7,072 entities / 1,826 relations & Full-text NER (4 types) and RE (11 types) in NLP papers & Micro-F\textsubscript{1}, Precision/Recall \\
& MASSW~\cite{zhang2025massw} & 2025 & 152K papers / 17 venues & Scientific workflow extraction across AI/ML venues (5 dimensions) & BLEURT, BERTScore, FActScore \\
& MIR-MultiCite~\cite{garikaparthi2025mir} & 2025 & 929 proposals / 10.2K cited papers & Methodology inspiration retrieval for CS research proposals & Recall@k, mAP \\
\midrule
\multirow{8}{*}{\centering\rotatebox[origin=c]{90}{\textbf{Understanding}}}
& ScienceQA~\cite{lu2022learn} & 2022 & 21K question-answer pairs & Multimodal science QA with reasoning chains across K--12 subjects & Accuracy \\
& ChartBench~\cite{xu2023chartbench} & 2023 & 600K question-answer pairs & Visual reasoning over scientific charts and figures & Accuracy, Confusion Rate \\
& KIWI~\cite{xu2024kiwi} & 2024 & 234 writing sessions & Knowledge-intensive writing grounded in multi-paper evidence & Human Rating, Win-Rate \\
& M3SciQA~\cite{li2024m3sciqa} & 2024 & 1,452 question-answer pairs & Cross-modal QA over multi-document scientific figures and text & MRR, Recall@k, nDCG@k \\
& SciAssess~\cite{cai2025sciassess} & 2025 & 6,888 questions / 27 tasks & Scientific literature analysis across chemistry subfields (27 tasks) & Accuracy, F1, Molecular Sim. \\
& MEBench~\cite{lin2025mebench} & 2025 & 4,780 question-answer pairs & Cross-document multi-entity reasoning in scientific papers & Accuracy, Entity-Aware F1 \\
& PaperArena~\cite{wang2025paperarena} & 2025 & 784 question-answer pairs & Tool-augmented agentic reasoning over scientific literature & Trajectory Acc., Efficiency \\
& ScholarQuery (Auto/Real)~\cite{he2025pasa} & 2025 & 35.5K synthetic / 50 real queries & Agentic paper search and synthesis over large academic corpora & Recall@k, Comprehensive Eval. \\
\midrule
\multirow{6}{*}{\centering\rotatebox[origin=c]{90}{\textbf{Application}}}
& HiCaD~\cite{feng2023hierarchical} & 2023 & 7,637 catalogues / 389K papers & Hierarchical catalogue generation for scientific literature & ROUGE, BERTScore, Edit Dist. \\
& SciReviewGen~\cite{kasanishi2023scireviewgen} & 2023 & 10K$+$ reviews / 690K papers & Automatic literature review generation from large paper corpora & ROUGE, BERTScore \\
& CHIME~\cite{hsu2024chime} & 2024 & 2,174 hierarchies / 472 topics & Hierarchical organization of biomedical studies for systematic review & Tree Precision/Recall \\
& ArxivDIGESTables~\cite{newman2024arxivdigestables} & 2024 & 2,228 tables / 7,542 papers & Cross-paper table synthesis and structured schema generation & Schema R/P, Value EM \\
& SurveyBench~\cite{yan2025surveyforge} & 2025 & 100 surveys / 10 topics & Multi-dimensional survey quality evaluation across 10 CS topics & Win-Rate, Human Ratings \\
& SurveyGen~\cite{bao2025surveygen} & 2025 & 4,205 surveys / 242,143 cited references & Quality-aware survey generation with structural consistency scoring & Structural Consistency Score \\
\bottomrule
\end{tabular}
\end{adjustbox}
\end{table*}

\paragraph{Scientific Knowledge Acquisition}
Acquisition benchmarks evaluate extraction fidelity across entities, relations, and larger workflow structures. PubTator 3.0, SciNLP, MASSW, and MIR-MultiCite together show the progression from closed-set biomedical extraction to full-paper workflow decomposition and graph-grounded retrieval~\cite{wei2024pubtator,duan2025scinlp,zhang2025massw,garikaparthi2025mir}. The main pattern is that narrowly defined extraction remains relatively tractable, whereas open-ended full-text structuring is still vulnerable to hallucination and schema inconsistency.

\paragraph{Scientific Content Understanding}
Understanding benchmarks span multimodal QA, chart reasoning, scientific writing, cross-document reasoning, and tool-augmented paper analysis~\cite{lu2022learn,xu2023chartbench,xu2024kiwi,li2024m3sciqa,cai2025sciassess,lin2025mebench,wang2025paperarena}. Two difficulties recur: sustained multi-step scientific reasoning remains weaker than surface comprehension, and multimodal scientific artifacts continue to trigger hallucination and grounding failures.

\paragraph{Scientific Knowledge Application}
Application benchmarks expose the sharpest mismatch between fluent generation and trustworthy scientific synthesis. Tasks such as hierarchical catalogue construction, literature review generation, table synthesis, and survey evaluation show that current models can produce polished outputs, but still struggle with logical organization, citation fidelity, and coherent long-form synthesis~\cite{feng2023hierarchical,kasanishi2023scireviewgen,hsu2024chime,newman2024arxivdigestables,yan2025surveyforge,bao2025surveygen}.

\paragraph{Progress, Limitations, and Future Directions}
Assistant benchmarks have matured across acquisition, understanding, and application, with broader multimodal coverage and more reproducible evaluation protocols~\cite{cai2025sciassess,yan2025surveyforge,wang2025paperarena}. Yet three limitations remain central: heavy reliance on closed-form outputs and lexical metrics, weak visibility into intermediate reasoning and citation-verification behavior, and benchmark concentration in computer science and biomedicine. Future work should emphasize open-ended scientific tasks, process-oriented evaluation, and broader cross-domain coverage.

\subsection{Observations}

The literature reviewed in this section suggests that LLMs can reduce the manual burden of information-intensive auxiliary work in scientific research, especially in retrieval, extraction, and first-pass synthesis. However, these gains do not extend uniformly across cognitive levels: open-ended application tasks still remain limited by citation errors, unsupported synthesis, and structural inconsistency. In practice, the Assistant role is therefore best characterized as shifting researcher effort from information gathering toward verification and interpretation rather than eliminating that effort altogether.

This pattern suggests that Assistant systems function primarily as tools for epistemic compression: they shorten the path from a large body of literature to a usable intermediate representation. Their main value therefore lies upstream in filtering, structuring, and organizing scientific material rather than downstream in making final epistemic judgments. This helps explain why benchmark improvements on extraction or question answering do not automatically translate into trustworthy survey writing or concept formation. Once a task requires not only locating evidence but deciding how it should be combined, weighted, and generalized, human expertise remains central to reliability.

\section{LLMs as Collaborators: Fostering Scientific Discovery}
\label{llmascoll}

The Collaborator role extends beyond research support to candidate-level scientific contribution. In this setting, LLMs are used not only to analyze existing knowledge, but also to propose hypotheses, experimental designs, and exploratory directions that remain subject to human grounding and validation.

\subsection{Role Definition}
\label{llm-collab-task}

\begin{figure*}[t]
    \centering
    \includegraphics[width=\linewidth]{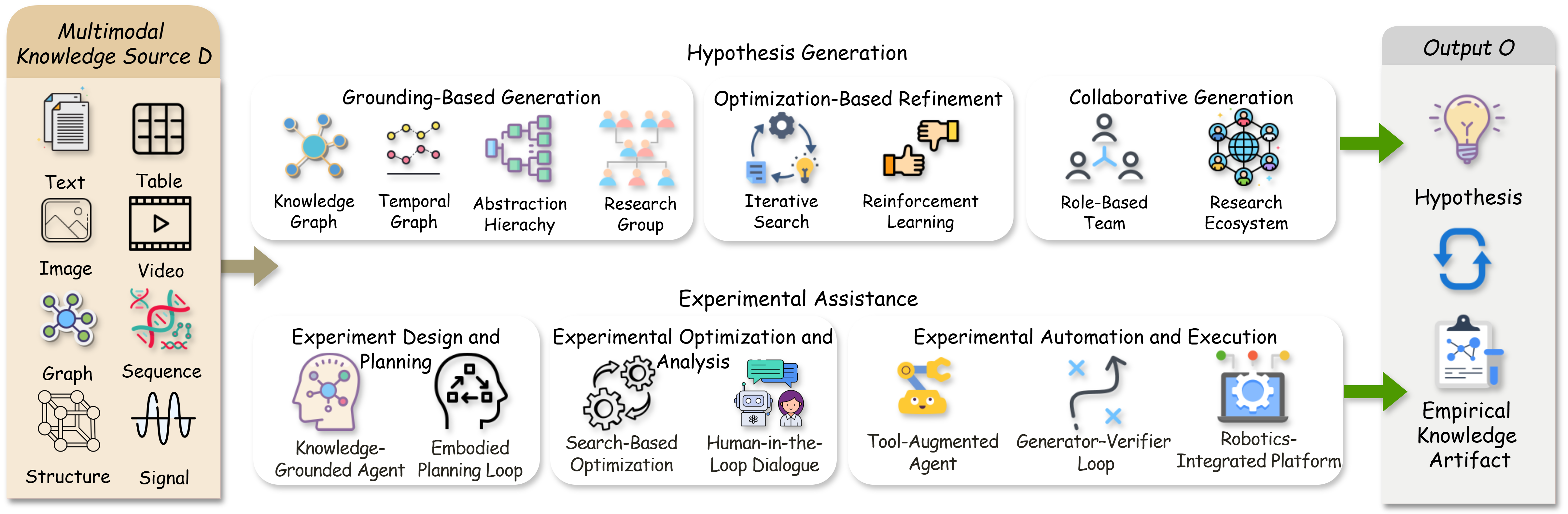}
    \caption{
        The framework of LLMs as collaborators for scientific discovery. Given a scientific query and multimodal knowledge sources, the collaborator supports hypothesis generation through grounding-based, optimization-based, and collaborative generation, and supports experimental assistance through experiment design and planning, experimental optimization and analysis, and experimental automation and execution.
    }
    \label{fig:collaborator-framework}
\end{figure*}

We formalize the LLM-as-Collaborator role through the following definition:
\begin{equation}
    \text{LLM}_{\text{Collaborator}}(Q,\, \mathcal{D},\, \theta) \;\rightarrow\; O \;\in\; \{\mathcal{H},\, \mathcal{E}\}
    \label{eq:collaborator}
\end{equation}
where $Q$, $\mathcal{D}$, and $\theta$ are as defined in Equation~\ref{eq:assistant}, and $O$ denotes either $\mathcal{H}$, a set of novel testable hypotheses produced at the \textit{Analyzing} and \textit{Evaluating} levels, or $\mathcal{E}$, empirical artifacts produced through experimental validation at the \textit{Creating} level. Unlike the terminal $O$ in Equation~\ref{eq:assistant}, $\mathcal{H}$ and $\mathcal{E}$ stand in a mutually conditioning relationship, where outcomes in $\mathcal{E}$ may confirm, refine, or falsify elements of $\mathcal{H}$, driving iterative cycles that progressively expand $\mathcal{D}$. Figure~\ref{fig:collaborator-framework} illustrates the two major functions of this role: hypothesis generation and experimental assistance.

\subsection{Tasks and Approaches}

\subsubsection{Hypothesis Generation}
\label{sec:hypothesis}

Hypothesis generation requires LLMs to operate at the \textit{Analyzing} and \textit{Evaluating} levels of Bloom's Taxonomy: decomposing existing knowledge in $\mathcal{D}$ to identify explanatory gaps, and assessing the plausibility of competing candidate hypotheses before committing to $H$. Existing work clusters into three methodological paradigms: grounding-based generation, optimization-based refinement, and collaborative generation.

\paragraph{(1) Grounding-Based Generation}
Grounding-based methods treat external knowledge as a first-class constraint on hypothesis space to reduce plausible but unsupported proposals. Early work relied mainly on domain-specific knowledge graphs and curated relations, especially in biomedicine~\cite{wang2019paperrobot,xiong2024improving}. Later systems broadened grounding to temporal concept graphs, abstraction hierarchies, and researcher-specific graphs that support cross-disciplinary transfer and alignment with prior work~\cite{xu2023exploring,gu2024llms,gu2024interesting}. In current pipelines, grounding serves both generative and evaluative functions: it shapes proposal space through retrieval or graph-conditioned reasoning, and is increasingly reused to assess novelty and faithfulness after generation~\cite{wang2024scimon,baek2025researchagent,radensky2024scideator}. The overall trend is a shift from using structured knowledge as a hard constraint toward integrating it throughout the full ideation loop.

\paragraph{(2) Optimization-Based Refinement}
Optimization-based frameworks treat hypothesis generation as iterative search rather than a single-shot mapping from $(Q, \mathcal{D}, \theta)$ to $H$. The MOOSE line illustrates this progression clearly, moving from multi-stage inspiration-driven ideation to chemistry-specific evolutionary search, hierarchical abstraction, and eventual incorporation of experimental feedback~\cite{yang2024large,yang2025moosechem,yang2025moosechem2,liu2025moose}. Parallel work generalizes the same idea through generate--evaluate--revise loops, UCB-based exploration, reward-shaped reinforcement learning, inference-time adversarial self-critique, and preference optimization~\cite{wang2023hypothesis,qiu2023phenomenal,zhou2024hypothesis,li2024learning,guo2025infal,dasguptaempowering}. Across these approaches, the central design question is how to balance broader exploration against tighter quality control when novelty, feasibility, and correctness cannot be optimized simultaneously.

\paragraph{(3) Collaborative Generation}
Collaborative generation approaches simulate the distributed cognition of scientific teams by assigning distinct roles to different agents and coordinating them through structured communication. Existing systems range from small role-based teams that separate literature analysis, method design, synthesis, and critique to larger virtual research ecosystems that simulate team formation, topic negotiation, and iterative evaluation in a shared environment~\cite{qi2024large,kumbhar2025hypothesis,su-etal-2025-many}. What these systems add over single-agent generation is not only more output diversity, but also more explicit internal critique, since different agents can be optimized for different subproblems such as literature coverage, feasibility checking, or adversarial review. Relative to single-agent baselines, they often improve diversity and critical coverage, but they also introduce new costs in coordination, token efficiency, and error propagation across long interaction traces.

\subsubsection{Experimental Assistance}
\label{sec:experiment}

Experimental assistance requires LLMs to operate at the \textit{Creating} level of Bloom's Taxonomy: translating hypotheses into executable protocols, navigating high-dimensional parameter spaces, and coordinating physical laboratory operations. The literature is organized into three functional phases that reflect the sequential logic of experimental workflow.

\paragraph{(1) Experiment Design and Planning}

A central finding in this literature is that general-purpose LLMs require substantial domain grounding to produce scientifically rigorous plans. One response is to couple LLM reasoning with structured domain knowledge through specialized multi-component agents, as in systems for gene editing, materials design, and scientific tool orchestration~\cite{huang2024crispr,kang2024chatmof,buehler2024generative,ding2025scitoolagent}. A complementary response grounds planning in literature-derived databases or multimodal scientific representations, including molecular structures, protein sequences, and 3D design spaces~\cite{zhang2024large,zhuang2025advancing,he2024novo,qi2025metascientist}. More recent work extends this pattern into embodied settings in which perception, robotics, and interactive guidance are integrated into a shared planning loop~\cite{cong2025labos}. One distinction across these systems is whether grounding enters as static prior knowledge or as live environmental information; the latter becomes increasingly important once planning must adapt to changing laboratory conditions rather than only propose protocols offline. Across these lines of work, the literature suggests that planning quality improves when LLMs are embedded in richer domain-specific knowledge and action environments.

\paragraph{(2) Experimental Optimization and Analysis}

Optimization systems must refine hypotheses and operating conditions in response to empirical feedback while balancing search breadth against evaluation cost. Existing approaches span uncertainty-guided Bayesian optimization, bilevel frameworks that separate conceptual search from simulator-based quantitative evaluation, and representation-level optimization through natural language feedback~\cite{ramos2023bayesian,sprueill2024chemreasoner,sprueill2023monte,ma2024llm,lv2025exploiting,yuksekgonul2025optimizing}. A contrasting line of work keeps human interaction central by using dialogue, retrieval, and expert feedback as the main optimization signal in tasks such as molecular refinement, multi-property drug design, comorbidity prediction, and AI experiment recommendation~\cite{liu2024conversational,ye2025drugassist,lu2024can,li2025agentexpt}. Taken together, these systems show that optimization quality depends not only on search strategy, but also on how tightly feedback is coupled to domain-specific evaluation.

\paragraph{(3) Experimental Automation and Execution}

Experimental automation requires bridging natural language instructions with physically executable laboratory operations. Existing systems cover modular chemistry agents, tool-augmented retrosynthesis assistants, generator--verifier loops for standardized action languages, and robotics-integrated platforms for electrochemistry and microscopy~\cite{boiko2023autonomous,m2024augmenting,yoshikawa2023large,ruan2024automatic,darvish2025organa,mandal2025evaluating}. Compared with planning-oriented systems, these execution-oriented agents expose sharper failure modes because an incorrect tool call or unsafe action sequence can have immediate physical consequences. Broader orchestration frameworks further show that explicit representations of tool capabilities can reduce failure rates in multidisciplinary settings~\cite{ding2025scitoolagent}. At the same time, this literature consistently surfaces the same limitations: safety filtering remains fragile, error propagation grows quickly across long action chains, and physical execution introduces reliability requirements that current agents only partially satisfy.

\subsection{Benchmarks and Evaluation}

Table~\ref{tab:collab_benchmarks} presents representative benchmarks for hypothesis generation and experimental assistance. These benchmarks reflect a progressive shift from surface-level lexical metrics toward multi-dimensional criteria encompassing novelty, truthfulness, feasibility, and end-to-end execution autonomy, as demanded by the generative and creative nature of the Collaborator role.

\begin{table*}[htbp]
\centering
\scriptsize
\setlength{\extrarowheight}{4pt}
\caption{Representative Benchmarks for LLMs as Collaborators.}
\label{tab:collab_benchmarks}
\begin{adjustbox}{max width=\textwidth}
\begin{tabular}{@{}m{0.5cm}p{2cm}cp{3.2cm}p{4.7cm}p{3.2cm}@{}}
\toprule
\textbf{Area} & \textbf{Benchmark} & \textbf{Year} & \textbf{Scale} & \textbf{Domain} & \textbf{Eval.\ Metrics} \\
\midrule
\multirow{9}{*}{\centering\rotatebox[origin=c]{90}{\textbf{Hypothesis Generation}}}
& SciMON~\cite{wang2024scimon}                 & 2024 & 67,408 NLP / 5,708 biomedical papers          & NLP and biomedical hypothesis generation with knowledge graph grounding          & Relevance, Utility, Novelty, Technical Depth \\
& TOMATO~\cite{yang2024large}                  & 2024 & 50 social science papers                      & Open-domain hypothesis discovery from social science literature                  & Validity, Novelty, Usefulness, Complexity Classification \\
& BioHypoGen~\cite{qi2024large}                & 2024 & 2,700 seen / 200 unseen pairs                 & Biomedical hypothesis generation with zero-shot generalization                   & Generalization, Zero-shot Hypothesis Generation \\
& LiveIdeaBench~\cite{Ruan2024LiveIdeaBenchEL} & 2024 & 1,180 scientific keywords / 18 domains        & Divergent idea generation from minimal context across 18 disciplines             & Originality, Feasibility, Fluency, Flexibility, Clarity \\
& TOMATO-Chem~\cite{yang2025moosechem}         & 2025 & 51 chemistry research papers                  & Chemistry hypothesis rediscovery against recent literature                       & Hit Ratio, Matched Score, Average Rank Ratio \\
& IdeaBench~\cite{guo2025ideabench}            & 2025 & 2,374 target / 29,408 reference papers        & Biomedical research idea generation with retrieval-based novelty scoring         & BERTScore, LLM Overlap, Insight Score \\
& TruthHypo~\cite{xiong2025truthhypo}          & 2025 & 2,024 instances / 3 relation tasks            & Truthful biomedical hypothesis generation across chemical, disease, and gene relations & Link-level F1, Relation-level Accuracy, KnowHD Groundedness Score \\
& ResearchBench~\cite{liu2025researchbench}    & 2025 & 1,386 research papers                         & Scientific discovery via inspiration retrieval and hypothesis composition         & Inspiration Retrieval, Hypothesis Composition, Ranking \\
& HypoBench~\cite{liu2025hypobench}            & 2025 & 194 datasets / 12 domains                     & Hypothesis quality evaluation across 7 real and 5 synthetic domains              & Explanatory Power, Plausibility, Interestingness \\
\midrule
\multirow{11}{*}{\centering\rotatebox[origin=c]{90}{\textbf{Experimental Assistance}}}
& BioPlanner~\cite{ODonoghue2023BioPlannerAE}  & 2023 & 100 biology protocols                         & Automated biology protocol planning and argument generation                      & Function Accuracy, Argument Precision \\
& MLAgentBench~\cite{huang2024mlagentbench}    & 2024 & 13 ML research tasks                          & Open-ended ML experimentation on real coding and modeling tasks                  & Success Rate \\
& SciCode~\cite{tian2024scicode}               & 2024 & 338 subproblems / 80 main problems            & Scientific coding across 16 natural science subfields                            & Problem Solve Rate \\
& TaskBench~\cite{shen2024taskbench}           & 2024 & 28,000+ tool-chain instances                  & Tool-chaining and task automation across APIs, ML models, and multimedia         & Node F1, Edge F1, Parameter Accuracy \\
& SUPER~\cite{bogin2024super}                  & 2024 & 801 problems across 3 sets                    & Research repository setup and execution from natural language instructions        & End-to-end \& Sub-task Success Rate \\
& MLE-bench~\cite{chan2025mle}                 & 2025 & 75 Kaggle competition tasks                   & End-to-end ML engineering on real Kaggle competitions                            & Medal Achievement Rate \\
& BoxingGym~\cite{gandhi2025boxinggym}         & 2025 & 10 probabilistic modeling domains             & Automated experimental design and probabilistic model discovery                  & Expected Information Gain (EIG) \\
& Curie~\cite{kon2025curie}                    & 2025 & 46 tasks / 4 CS domains                      & AI-driven scientific experiment planning, execution, and interpretability         & Experimental Rigor Score (Intra/Inter-Agent, Interpretability) \\
& MLGym~\cite{nathani2025mlgym}               & 2025 & 13 AI research tasks                          & AI research agents on data science, vision, RL, and game theory tasks            & AUP Score, Baseline Improvement Rate \\
& PaperBench~\cite{starace2025paperbench}      & 2025 & 20 ICML 2024 Spotlight papers / 8,316 tasks   & AI replication of ICML papers with hierarchical rubric scoring                   & Hierarchical Rubric Replication Score \\
& EXP-Bench~\cite{kon2026expbench}             & 2026 & 461 tasks / 12,737 subtasks from 51 papers    & AI experiment conduction from NeurIPS/ICLR 2024 papers across RL, AI applications, and generative models & Design, Implementation, Execution, Conclusion Correctness; All-correct Rate \\
\bottomrule
\end{tabular}
\end{adjustbox}
\end{table*}

\paragraph{Hypothesis Generation}
Hypothesis-generation benchmarks now span three complementary emphases: large-scale automatic scoring, expert-annotated domain evaluation, and reliability-oriented testing. SciMON and IdeaBench emphasize scale but depend heavily on automatic metrics~\cite{wang2024scimon,guo2025ideabench}; TOMATO, TOMATO-Chem, and ResearchBench provide stronger domain-sensitive evaluation through expert annotation and structured subtask design~\cite{yang2024large,yang2025moosechem,liu2025researchbench}; and BioHypoGen, HypoBench, LiveIdeaBench, and TruthHypo focus more directly on generalization, creativity, and factual grounding~\cite{qi2024large,liu2025hypobench,Ruan2024LiveIdeaBenchEL,xiong2025truthhypo}. Across these benchmarks, the main pattern is not uniform progress but a persistent trade-off among novelty, feasibility, and factual reliability.

\paragraph{Experimental Assistance}
Experimental-assistance benchmarks likewise range from narrow subtask evaluation to end-to-end research execution. MLAgentBench, MLE-bench, TaskBench, and SciCode probe controlled experimentation, orchestration, and scientific coding~\cite{huang2024mlagentbench,chan2025mle,shen2024taskbench,tian2024scicode}, whereas SUPER, PaperBench, Curie, and EXP-Bench test increasingly complete research pipelines~\cite{bogin2024super,starace2025paperbench,kon2025curie,kon2026expbench}. Cross-disciplinary settings such as BioPlanner, BoxingGym, and MLGym broaden the scope beyond mainstream ML evaluation~\cite{ODonoghue2023BioPlannerAE,gandhi2025boxinggym,nathani2025mlgym}. The shared conclusion is that strong subtask performance does not yet translate into reliable end-to-end experimental autonomy.

\paragraph{Progress, Limitations, and Future Directions}
The Collaborator benchmark ecosystem has matured substantially, but three limitations remain central. First, many evaluations still rely on metrics that are only weakly aligned with scientific value. Second, benchmark designs rarely connect generated hypotheses to downstream validation, leaving the novelty-feasibility-truthfulness tension unresolved. Third, subtask success continues to overstate true autonomy because failures compound sharply across full research pipelines. Future work should prioritize longitudinal evaluation, stronger grounding and truthfulness measures, and benchmark designs that connect ideation to experimental validation~\cite{liu2025hypobench,xiong2025truthhypo,kon2026expbench}.

\subsection{Observations}

Evidence across Collaborator-role benchmarks suggests that LLMs expand the space of hypotheses, designs, and experimental options available for scientific exploration, but do not yet function as reliable generators of validated scientific knowledge. Across both hypothesis generation and experimental assistance, the recurring pattern is productive expansion under continued human supervision, with limitations arising from trade-offs among novelty, feasibility, factual grounding, and pipeline reliability~\cite{liu2025hypobench,xiong2025truthhypo,kon2026expbench}. Within the role taxonomy, this means that collaborator systems contribute candidate scientific value, but that value remains conditional on downstream validation. The Collaborator role is therefore most plausibly understood, at present, as a human-supervised aid to scientific exploration.

One implication of these findings is that the central bottleneck of the Collaborator role lies less in idea production than in the transition from candidate idea to accountable scientific claim. LLMs can already generate a broad set of plausible directions, yet scientific value depends on whether those directions survive grounded scrutiny, experimental follow-up, and selective allocation of time and resources. Collaborator systems therefore appear most useful when breadth matters more than immediate correctness, for example in early-stage exploration or when researchers wish to probe underexamined regions of a search space.

\section{LLMs as Scientists: Autonomous Research and Discovery}
\label{llmasauto}

The Scientist role refers to higher-autonomy settings in which LLM-based systems attempt to execute larger portions of the research loop with limited step-by-step human intervention~\cite{lu2026ai,weng2025cycleresearcher,wang2023scientific}. The central question in this section is not whether such systems are fully independent in practice, but how close current architectures come to closed-loop research and discovery under benchmarked conditions.

\subsection{Role Definition}
\label{llm-scientist-task}
\begin{figure*}[t]
    \centering
    \includegraphics[width=0.9\linewidth]{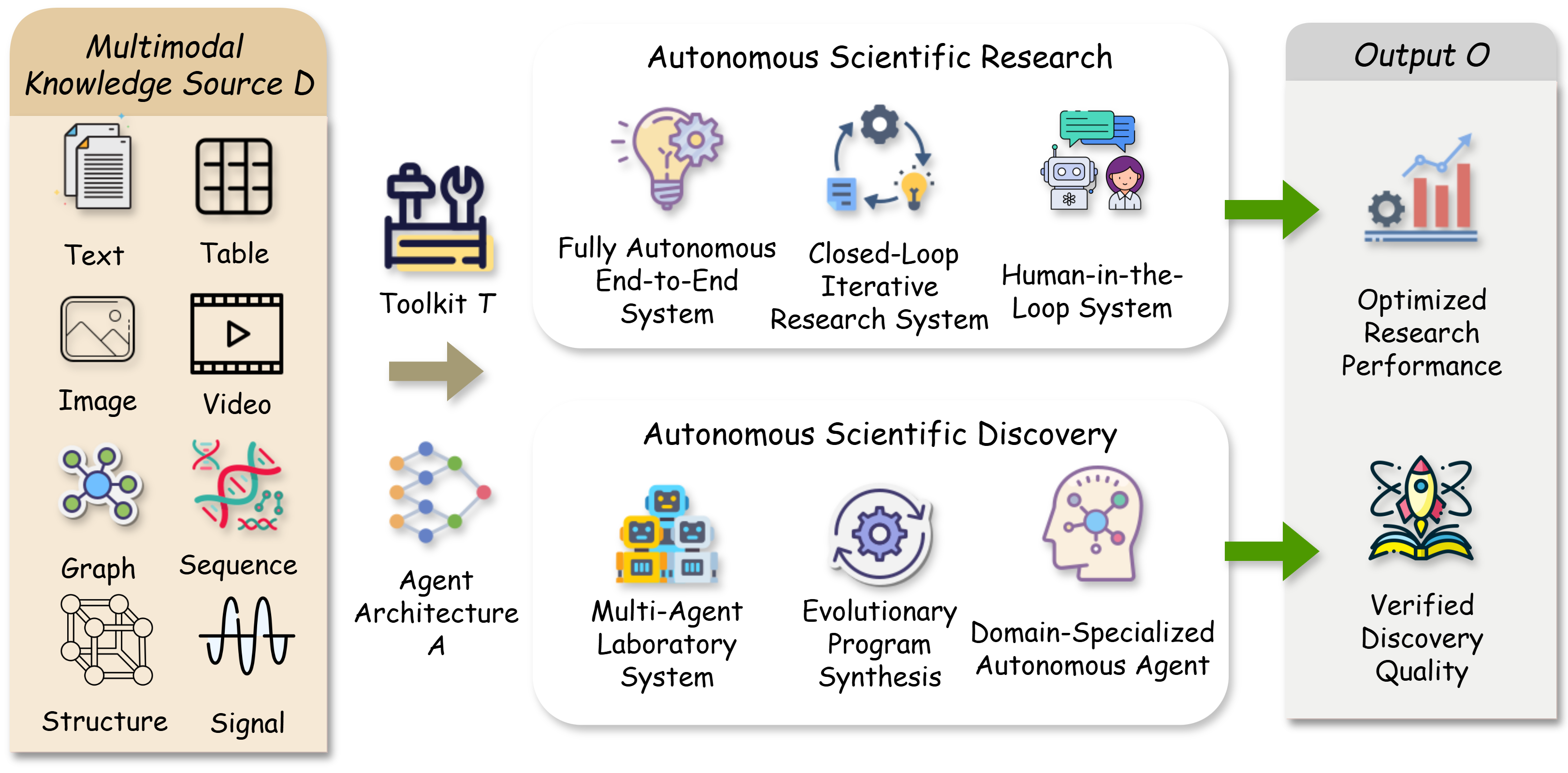}
    \caption{
        The framework of LLMs as scientists for autonomous research and discovery. The system combines multimodal knowledge, agent architecture, and toolkits to support autonomous research workflows and autonomous discovery pipelines, producing optimized research performance and verified discovery quality.
    }
    \label{fig:scientist-framework}
\end{figure*}

We formalize the LLM-as-Scientist role through the following definition:

\begin{equation}
    \text{LLM}_{\text{Scientist}}(Q,\, \mathcal{D},\, \theta,\, \mathcal{A},\, \mathcal{T}) \;\rightarrow\; O \;\in\; \{\mathcal{SR}^{*},\, \mathcal{SD}^{*}\}
    \label{eq:scientist}
\end{equation}
where $Q$, $\mathcal{D}$, and $\theta$ are as defined in Equation~\ref{eq:assistant}, $\mathcal{A}$ is the agent architecture governing task decomposition and coordination, and $\mathcal{T}$ the toolkit of external tools, simulators, and verifiers. The output $O$ is typed by the system's mandate: $\mathcal{SR}^{*}$ denotes peak process efficiency achieved by optimizing over $\mathcal{A}$ and $\mathcal{T}$, and $\mathcal{SD}^{*}$ denotes novel discoveries produced through iterative \textsc{Plan}$\rightarrow$\textsc{Act}$\rightarrow$\textsc{Observe}$\rightarrow$\textsc{Refine} cycles until a convergence or budget criterion is met. Figure~\ref{fig:scientist-framework} illustrates how LLM scientists integrate multimodal knowledge, agent architecture, and external toolkits to support both autonomous scientific research and discovery.

\subsection{Tasks and Approaches}

\subsubsection{Autonomous Scientific Research}

Autonomous scientific research asks whether an LLM-based system can independently execute the full research workflow, from problem formulation and experimental design through implementation, analysis, and manuscript production, with outputs evaluated on methodological soundness and pipeline efficiency. Systems in this space are organized by the degree of human oversight: fully autonomous end-to-end pipelines, closed-loop iterative frameworks that couple reasoning with execution feedback, and human-in-the-loop architectures that admit expert intervention at defined checkpoints.

\paragraph{(1) Fully Autonomous End-to-End Systems}
Fully autonomous pipelines delegate most stages of the scientific method to a single agent or tightly coupled agent pair. Representative systems differ mainly in how they structure iteration and verification: some pair research generation with automated review, some widen exploration through tree search or mentor-provided baselines, and others emphasize traceability by decomposing the process into deterministic steps or by training general analyst-style agents through curriculum learning~\cite{weng2025cycleresearcher,lu2026ai,yamada2025ai,miyai2025jr,ifargan2025autonomous,zhang2025deepanalyze}. They also differ in target artifact: some aim to produce full paper-like outputs, whereas others prioritize auditable reports or bounded improvements over an existing baseline. The shared objective is end-to-end pipeline closure, while the main design variation lies in how aggressively systems trade open-ended search for control and auditability.

\paragraph{(2) Closed-Loop Iterative Research Systems}
Closed-loop frameworks couple hypothesis generation to executable feedback, treating research and discovery as optimization problems whose reward derives from observed outcomes. Existing systems instantiate this pattern through prompt-level recycling of execution results, compile--run--debug loops, self-evolving idea modules with exception-guided repair, evolutionary search over code or paper fragments, and dynamic task graphs that are rewired as intermediate results accumulate~\cite{yuan2025dolphin,li2024mlr,team2025internagent,jansen2025codescientist,hu2025flowsearch}. Compared with end-to-end pipelines, these systems usually place less emphasis on polished final artifacts and more emphasis on repeated correction from intermediate failures, which makes them a useful testbed for studying error recovery. Supporting infrastructures such as ToolUniverse broaden the set of tools that can be invoked within these loops, but do not remove the core challenge of maintaining coherent long-horizon strategy across many dependent steps~\cite{gao2025democratizing}.

\paragraph{(3) Human-in-the-Loop Systems}
Human-in-the-loop frameworks balance agent autonomy with structured expert intervention. Representative systems vary in how they expose intervention points: some inject guidance at each stage of a role-based research pipeline, some allow autonomous laboratories to build cumulatively on shared prior reports, and others combine dynamic workflow generation with memory persistence and non-blocking human feedback~\cite{schmidgall-etal-2025-agent,schmidgall2025agentrxiv,li2025build}. These designs also make clear that oversight need not be continuous to matter; sparse intervention at high-leverage checkpoints can substantially redirect search trajectories or prevent low-value execution. SciSciGPT makes this design space explicit by pairing multi-agent workflow automation with a capability maturity model for human-AI collaboration~\cite{shao2025sciscigpt}. Taken together, these systems suggest that partial oversight remains valuable not only for safety, but also for steering open-ended search.

\subsubsection{Autonomous Scientific Discovery}
\label{llm-scientist-discovery}

Autonomous scientific discovery asks whether an LLM-based system can produce knowledge that is genuinely new to humanity, such as novel materials, drug candidates, or mathematical theorems, and that withstands external verification. Where research systems are evaluated on process correctness, discovery systems are evaluated on whether their outputs advance the frontier of what is known. Systems in this space are organized by algorithmic architecture: multi-agent laboratories built around role specialization and iterative peer critique, evolutionary program synthesis pipelines that enforce a strict generation-verification separation, and domain-specialized autonomous agents that embed scientific priors as pluggable architectural components.

\paragraph{(1) Multi-Agent Laboratory Systems}
The multi-agent laboratory paradigm decomposes the scientific process into distinct epistemic roles, such as a Planner, a Critic, an Experimenter, and a Coder, which interact through iterative cycles of hypothesis generation, experimental execution, and peer critique~\cite{liu2024aigs,jiang2025agenticsciml}. \textcite{boiko2023autonomous} provided an early and influential realization of this paradigm, demonstrating that LLM modules specialized respectively for web search, technical documentation retrieval, code execution, and robotic experimentation APIs could be orchestrated to autonomously design and perform complex chemical syntheses. This role-based architecture attenuates hallucinations through iterative peer review and naturally accommodates heterogeneous external tools, since each agent can be assigned a specialized toolbox without disrupting the overall coordination protocol.

Several systems extend this paradigm in ways intended to improve rigor. At the hypothesis level, some frameworks enforce falsifiability by dedicating a FalsificationAgent to designing ablation experiments that actively test candidate hypotheses~\cite{liu2024aigs}, while others ground inter-agent debate in structured knowledge representations such as knowledge graphs~\cite{ghafarollahi2024sciagents}. At the execution level, domain-adapted variants couple surrogate model scoring with self-reflection modules that evaluate candidate outcomes before issuing the next modification~\cite{jia2024llmatdesign}, and Planner--Instructor frameworks automate ML-based computational pipelines across drug discovery tasks~\cite{liu2024drugagent}. \textcite{jiang2025agenticsciml} further report that structured debate among specialized agents with retrieval-augmented memory and evolutionary search can outperform both single-agent and human-designed baselines in their benchmark setting. Taken together, these results suggest that agent ensembles may support broader search and stronger error correction than single-agent configurations, although the generality of this effect remains to be established across domains.

\paragraph{(2) Evolutionary Program Synthesis}
Where multi-agent systems emphasize collaborative reasoning, evolutionary program synthesis treats the LLM as a generative search operator whose outputs are executable artifacts, such as functions and equations, probabilistic programs, or entire codebases, submitted to a deterministic external evaluator for scoring~\cite{romera2024mathematical,shojaee2025llmsr,li2024automated,alphaevolve2025}. This strict generation-verification separation guarantees falsifiability and provides a principled objective for iterative refinement.

The core algorithmic pattern, introduced by \textcite{romera2024mathematical}, maintains a population of candidate programs, selects high-scoring individuals as seeds for the next generation, and preserves diversity through explicit mechanisms such as inter-island cloning or clustering. Later work refines both generation and evaluation: memory buffers and reflection modules bias sampling toward structurally novel but performant candidates, while composite scoring and probabilistic inference favor parsimonious, empirically plausible programs~\cite{shojaee2025llmsr,hu2026multif,li2024automated}. \textcite{alphaevolve2025} suggest that the same pattern can scale to industrial code optimization and mathematically difficult search spaces. Across these systems, the recurring commitments are persistent memory, explicit diversity preservation, and executable outputs that can be audited externally.

\paragraph{(3) Domain-Specialized Autonomous Agents}
A third paradigm embeds the architectural patterns above into specific scientific disciplines by treating domain priors as first-class modular components rather than external context injected through prompts~\cite{liu2025expert,gao2025txagent,xiao2026cellagent,wang2025spatialagent}. Curated tool libraries, simulation surrogates, and domain ontologies are packaged as interchangeable modules, enabling the agent to apply rigorous discipline-specific evaluation criteria at every step of the discovery loop~\cite{gao2025txagent,liu2025expert,wang2025spatialagent}.

This architectural choice yields measurable reliability gains across domains. Materials and drug-discovery systems close the loop by combining generation with feasibility checks, surrogate-model scoring, and multi-criteria screening~\cite{liu2025expert,averly2025liddia}. In biomedical, single-cell, and spatial reasoning, performance improves when agents can select from curated tool universes and maintain domain-specific planning and memory structures~\cite{gao2025txagent,xiao2026cellagent,wang2025spatialagent}. The broader implication is that the primary bottleneck in autonomous discovery is not language generation in isolation, but the fidelity with which domain knowledge is encoded into evaluation, memory, and tool-selection infrastructure. The cost of this strategy is reduced transferability: systems optimized for one domain typically require substantial redesign before they generalize to another.

\subsection{Benchmarks and Evaluation}

Table~\ref{tab:scientist-benchmarks} consolidates representative benchmarks for the LLM-as-Scientist role along two evaluative axes: autonomous scientific research, which measures pipeline completeness and process correctness, and autonomous scientific discovery, which assesses whether agent outputs genuinely advance the knowledge frontier.

\begin{table*}[htbp]
\centering
\scriptsize
\setlength{\extrarowheight}{4pt}
\caption{Representative Benchmarks for LLMs as Scientists.}
\label{tab:scientist-benchmarks}
\begin{adjustbox}{max width=\textwidth}
\begin{tabular}{@{}m{0.5cm}p{2cm}cp{3.0cm}p{4.7cm}p{3.5cm}@{}}
\toprule
\textbf{Area} & \textbf{Benchmark} & \textbf{Year} & \textbf{Scale} & \textbf{Domain} & \textbf{Eval.\ Metrics} \\
\midrule
\multirow{9}{*}{\rotatebox[origin=c]{90}{\parbox{6.5cm}{\centering\textbf{Autonomous Scientific Research}}}}
& LAB-Bench~\cite{laurent2024lab}             & 2024 & 2,400+ MC questions / 41 cloning scenarios    & Practical biology research tasks across 8 categories including literature, protocols, and sequences & Multiple-Choice Accuracy; Cloning Task Success Rate \\
& DSBench~\cite{jing2025dsbench}              & 2025 & 466 analysis / 74 modelling tasks             & Data science agent evaluation on realistic Kaggle and Eloquence tasks                              & Task Completion Rate; Prediction Accuracy \\
& CS-Bench~\cite{song2025cs}                  & 2025 & $\sim$10,000 items / 26 subfields             & Bilingual CS knowledge and reasoning evaluation across 4 CS areas                                  & Knowledge Accuracy; Reasoning Accuracy; Multilingual Performance \\
& AAAR-1.0~\cite{lou2025aaar}                 & 2025 & 1,049 equations / 993 papers / 11,376 review segments & Expert-level AI research tasks: equation inference, experiment design, weakness detection, review critique & Equation Inference F\textsubscript{1}; Experiment Design Quality; Weakness Detection F\textsubscript{1}; Review Critique Score \\
& Scientist-Bench~\cite{tang2025airesearcher} & 2025 & 28 tasks from 22 papers                       & End-to-end autonomous CS research from idea to manuscript                                          & Code Completeness; Manuscript Quality (LLM Peer Review Score) \\
& RE-Bench~\cite{wijk2025rebench}             & 2025 & 7 environments / 71 attempts by 61 human experts & ML research engineering with direct human expert comparison                                     & Normalized Score; Human vs.\ AI Score; Best-of-$k$ Performance \\
& AstaBench~\cite{bragg2025astabench}         & 2025 & 2,400+ problems / 11 benchmarks / 57 agents   & Holistic scientific research assistance across literature, code, data analysis, and discovery       & Per-task Accuracy; End-to-End Completion Rate; Cost-normalized Score \\
& ScienceBoard~\cite{sun2026scienceboard}     & 2026 & 169 tasks / 6 domains                         & Multimodal autonomous agents on realistic scientific workflows with professional software           & Overall Task Success Rate; Sub-task Success Rate \\
& InnovatorBench~\cite{wu2026innovatorbench}  & 2026 & 20 tasks / 6 LLM research categories          & End-to-end LLM research spanning data construction, loss design, and scaffold construction         & Reference-Surpass Rate; Time-to-Best-Performance \\
\midrule
\multirow{6}{*}{\rotatebox[origin=c]{90}{\parbox{5cm}{\centering\textbf{Autonomous Scientific Discovery}}}}
& BrainBench~\cite{luo2024large}                     & 2024 & 300 paired abstracts / 5 neuroscience subfields & Outcome prediction from neuroscience abstracts with human expert comparison                      & Outcome Prediction Accuracy; Human Expert Comparison \\
& DiscoveryWorld~\cite{jansen2024discoveryworld}     & 2024 & 120 tasks / 8 themes / 3 difficulty levels    & End-to-end scientific discovery in simulated environments across diverse topics                    & Task Completion; Procedural Relevance; Explanatory Quality \\
& ScienceAgentBench~\cite{chen2025scienceagentbench} & 2025 & 102 tasks from 44 papers                      & Data-driven discovery in bioinformatics, computational chemistry, GIS, and cognitive neuroscience  & Valid Execution Rate; Success Rate; CodeBERTScore; API Cost \\
& DiscoveryBench~\cite{majumder2025discoverybench}   & 2025 & 264 real-world / 903 synthetic tasks          & Data-driven hypothesis discovery across sociology, biology, economics, and engineering             & Hypothesis Accuracy; Facet-level Evaluation (Context, Variable, Relation) \\
& CellBench~\cite{alber2025cellvoyager}              & 2025 & 76 scRNA-seq studies / 483 analyses           & Autonomous computational biology analysis over single-cell RNA sequencing datasets                 & Micro/Macro-averaged Workflow Reconstruction Accuracy \\
& NewtonBench~\cite{zheng2026newtonbench}            & 2026 & 324 tasks / 12 physics domains                & Generalizable physical law discovery via counterfactual law shifts                                & Law Discovery Accuracy; Exploration Efficiency; Counterfactual Generalization \\
\bottomrule
\end{tabular}
\end{adjustbox}
\end{table*}

\subsubsection{Autonomous Scientific Research}

Research benchmarks span a continuum from sub-skill diagnostics to full workflow execution. LAB-Bench, CS-Bench, DSBench, and AAAR-1.0 isolate specific competencies such as biological protocol reasoning, computer science knowledge, data analysis, experiment design, and review critique~\cite{laurent2024lab,song2025cs,jing2025dsbench,lou2025aaar}. Scientist-Bench, ScienceBoard, and InnovatorBench move closer to full research execution by evaluating hypothesis-to-manuscript workflows, tool-mediated scientific software use, and long-horizon innovation tasks~\cite{tang2025airesearcher,sun2026scienceboard,wu2026innovatorbench}. RE-Bench and AstaBench provide the strongest calibration of research autonomy by comparing agents against human experts or full research-chain completion requirements~\cite{wijk2025rebench,bragg2025astabench}. Taken together, these benchmarks show that current systems can perform many research subroutines well, but their reliability degrades sharply as workflow depth and task duration increase.

\subsubsection{Autonomous Scientific Discovery}

Discovery benchmarks impose a stricter criterion: outputs must be not only procedurally correct but also novel or externally verifiable. BrainBench and ScienceAgentBench evaluate discovery-relevant performance on fixed corpora and expert-designed tasks~\cite{luo2024large,chen2025scienceagentbench}; DiscoveryWorld and NewtonBench test exploratory behavior in interactive settings designed to reduce memorization effects~\cite{jansen2024discoveryworld,zheng2026newtonbench}; and DiscoveryBench and CellBench probe multi-step discovery over heterogeneous inputs~\cite{majumder2025discoverybench,alber2025cellvoyager}. Across these settings, domain knowledge and task structure matter greatly, but robust exploratory reasoning remains inconsistent.

\subsubsection{Progress, Limitations, and Future Directions}

Across research and discovery benchmarks, three patterns recur. First, component-level success consistently overstates end-to-end autonomy because errors compound across multi-step pipelines. Second, better tools do not by themselves solve long-horizon scientific reasoning; exploration strategy and error recovery remain central bottlenecks. Third, evaluation remains uneven: matched human baselines, memorization-resistant design, and explicit safety assessment are still rare. Future benchmark design should therefore prioritize longitudinal protocols, broader human baselines, memorization-resistant task construction, and multi-objective evaluation that includes reliability, novelty, reproducibility, and safety~\cite{wijk2025rebench,kon2026expbench,zheng2026newtonbench,bragg2025astabench}.

\subsection{Observations}

LLMs in the Scientist role have made the strongest move toward closing the research loop, but current evidence supports a qualified rather than absolute interpretation of this progress. Existing systems can autonomously complete substantial portions of problem formulation, experimentation, analysis, and reporting in constrained settings, yet benchmark results consistently show that reliability deteriorates as pipelines lengthen and tasks become more open-ended~\cite{wijk2025rebench,kon2026expbench}. The main bottlenecks are compounding execution failures, weak long-horizon strategy, and safety risks in tool-mediated or physical settings. In practice, the Scientist role is presently most credible where domain priors are tightly encoded, evaluation criteria are externally verifiable, and human oversight remains structurally embedded.

One distinction that emerges in this literature is between closing a workflow and closing an epistemic loop. Many systems can now chain together planning, execution, analysis, and reporting, but stronger forms of scientific autonomy require that the resulting conclusions remain robust under external verification and transfer beyond the benchmark setting. Current Scientist-role systems therefore appear stronger as autonomous research operators than as autonomous discoverers. The difference is important because pipeline completeness is primarily a systems property, whereas scientific credibility is a joint property of process, evidence, and outcome.

\section{LLMs as Evaluators: Validating Scientific Innovation}
\label{llmasval}

Where the preceding three roles are defined by their capacity to produce scientific knowledge artifacts, the Evaluator role is defined by the orthogonal capacity to validate them. An LLM operating as an Evaluator adjudicates the quality, validity, and significance of outputs generated by researchers or autonomous systems, functioning as a meta-level participant whose purpose is epistemic governance rather than epistemic generation.

\subsection{Role Definition}
\label{llm-eval-task}

\begin{figure*}[t]
    \centering
    \includegraphics[width=0.9\linewidth]{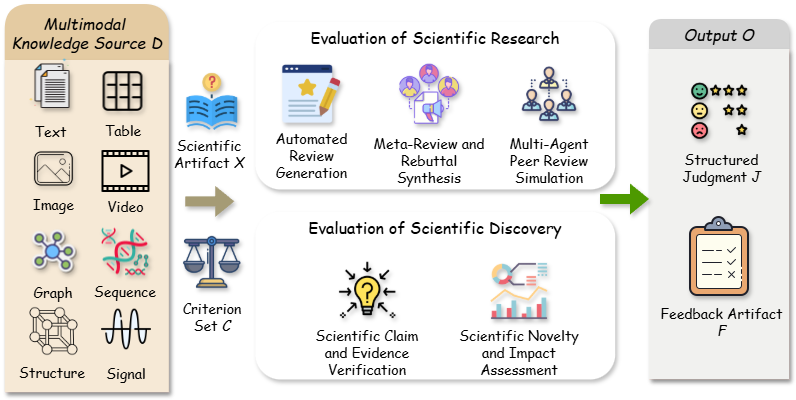}
    \caption{
        The framework of LLMs as evaluators for validating scientific innovation. The evaluator uses multimodal knowledge, scientific artifacts, and evaluation criteria to assess scientific research and scientific discovery, producing structured judgments and feedback artifacts.
    }
    \label{fig:evaluator-framework}
\end{figure*}

We formalize the LLM-as-Evaluator role through the following definition:

\begin{equation}
    \text{LLM}_{\text{Evaluator}}(\mathcal{X},\, \mathcal{C},\, \mathcal{D},\, \theta) \;\rightarrow\; O \;\in\; \{\mathcal{J},\, \mathcal{F}\}
    \label{eq:evaluator}
\end{equation}
where $\mathcal{X}$ is the artifact under evaluation (a manuscript, hypothesis, or empirical claim), $\mathcal{C}$ the criterion set against which $\mathcal{X}$ is assessed, $\mathcal{D}$ the background scientific knowledge base, and $\theta$ the model parameters governing evaluative reasoning. The output $O$ is typed by evaluative function: $\mathcal{J}$ denotes a structured judgment (a score, accept/reject decision, or validity rating), and $\mathcal{F}$ a feedback artifact (a structured critique or revision recommendation) that renders the reasoning behind $\mathcal{J}$ auditable. Figure~\ref{fig:evaluator-framework} illustrates how LLM evaluators assess scientific artifacts under explicit criteria through both research evaluation and discovery evaluation.

\subsection{Evaluation of Scientific Research}
\label{sec:eval-research}

Evaluation of scientific research corresponds to the setting in Equation~\ref{eq:evaluator} where $\mathcal{X}$ is a manuscript or peer review artifact. The central question is whether LLMs can genuinely assess the quality and value of scientific research, and peer review serves as the natural testbed: it is the institutionalized process through which the community judges whether a contribution is rigorous, novel, and significant enough to enter the scientific record~\cite{zhuang2025large}. An LLM producing $(J, F)$ pairs calibrated to expert reviewer judgments demonstrates not merely surface reading competence but a deeper capacity to reason about scientific credibility. Work in this area clusters into three directions: automated review generation, meta-review and rebuttal synthesis, and multi-agent peer review simulation.

\paragraph{(1) Automated Review Generation}
The foundational challenge is defined by \textcite{yuan2022can} who show that automatic review systems can cover more paper aspects than human reviewers while remaining less constructive and less factually grounded, and who identify eight open challenges for future systems. Many of these limitations persist in LLM-based review generation~\cite{li2025unveiling}. Subsequent work has pursued three main directions: stronger grounding through knowledge-guided generation~\cite{yuan2022kid}, deeper review reasoning through reflective or decompositional frameworks~\cite{zhu-etal-2025-deepreview,chang2025treereview}, and better alignment with expert signals through reinforcement learning or rebuttal supervision~\cite{zeng-etal-2025-reviewrl,wu2026rbtact}. LMCBert complements these efforts by targeting calibrated paper scoring rather than review text generation~\cite{liu2025lmcbert}. Despite measurable gains in coverage and fluency, current systems remain overly positive, insufficiently specific on methodological flaws, and weakly calibrated on novelty~\cite{zhou2024llm,du2024llms}.

\paragraph{(2) Meta-Review and Rebuttal Synthesis}
Where single-paper review generation maps one manuscript to one critique, meta-review generation must aggregate and reconcile multiple conflicting reviewer perspectives into a coherent editorial verdict. Early approaches model this as multi-document summarization with conversational structure~\cite{li2023summarizing, wu2022incorporating}, treating reviewer exchanges and author rebuttals as structured dialogue to be distilled. A persistent failure mode is that straightforward aggregation suppresses minority views: \textcite{chen2025bridging} address this through conflict-aware meta-review generation grounded in cognitive alignment principles from social psychology, explicitly modeling reviewer disagreement before synthesis. The $\mathrm{Re}^2$ dataset~\cite{zhang2025re} provides consistency-ensured annotations spanning initial reviews, author rebuttals, and multi-turn reviewer discussions, enabling training and evaluation of systems that must track position changes across the entire review lifecycle. The availability of rebuttal-aware training data and conflict-aware modeling opens a path toward improving the specificity and balance of generated meta-reviews, though current systems still struggle to faithfully represent minority reviewer positions and to produce verdicts whose reasoning chains are transparent enough for editorial accountability.

\paragraph{(3) Multi-Agent Peer Review Simulation}
A third direction treats peer review not as a generation task but as a sociotechnical process to be modeled and optimized through agent-based simulation. Existing systems vary in emphasis: some analyze how reviewer, author, and area-chair roles interact to produce collective bias~\cite{Jin2024AgentReviewEP}; some ground reviewer agents in retrieved prior literature to improve specificity~\cite{goyal2026scholarpeer}; some focus on explainable scoring pipelines or closed evaluation--revision loops~\cite{huang2025papereval,zhao2026apres}; and some study real-world review-feedback deployment at scale~\cite{thakkar2026large}. A useful contribution of this line of work is that it shifts attention from isolated review quality to review dynamics, including disagreement, influence, and revision across rounds. Taken together, this literature suggests that role decomposition improves coverage and consistency relative to single-model baselines, but also reproduces social influence effects that complicate calibration and accountability.

\paragraph{Progress, Limitations, and Future Directions}
Across all three directions, LLMs have demonstrated a credible capacity to replicate surface-level reviewer behavior: generating structured critiques, aggregating multiple perspectives into meta-reviews, and reproducing the distributional statistics of accept/reject decisions at above-chance rates. However, a consistent gap between surface competence and genuine evaluative judgment persists. LLM reviewers are systematically over-lenient, under-specific on methodological flaws, and poorly calibrated on novelty, the dimension most consequential for editorial decisions. Multi-agent simulation exposes an additional concern: the same social influence dynamics that distort human peer review are reproduced and potentially amplified in agent ensembles, raising the risk that LLM-assisted review inherits rather than corrects the biases of the process it is meant to improve. Future work should prioritize calibration against longitudinal expert judgments rather than single-round annotations, develop evaluation protocols that specifically target novelty and methodological soundness assessment, and address the accountability gap by ensuring that generated $(J, F)$ pairs are traceable to specific paper claims rather than aggregated impressions.

\subsection{Evaluation of Scientific Discovery}
\label{sec:eval-discovery}
 
Evaluation of scientific discovery concerns settings in which the artifact under assessment is a scientific claim, empirical finding, or candidate contribution. Relative to peer review of manuscripts, this setting places greater emphasis on evidence grounding, external validity, and prospective significance. The literature in this area is reviewed in two parts: scientific claim and evidence verification, and scientific novelty or impact assessment.

\paragraph{(1) Scientific Claim and Evidence Verification}
Scientific verification asks whether models can correctly map claims to supporting or refuting evidence. Work in this area has progressed from expert-annotated biomedical and public-health datasets to open-domain scientific claim verification, compositional table reasoning, and adversarial settings involving misrepresented scientific evidence~\cite{wadden2020fact,kotonya2020explainable,wadden2022scifact,lu2023scitab,glockner2024missci,glockner2025grounding}. Modeling advances such as zero-shot claim generation, weak supervision with full-document context, adversarial training, and modular evaluation frameworks have improved robustness, but performance still drops sharply under open-domain retrieval, dispersed evidence, and adversarial misrepresentation~\cite{wright2022generating,wadden2022multivers,karisani2024fact,iqbal2024openfactcheck,javaji2025can}.

\paragraph{(2) Scientific Novelty and Impact Assessment}
A separate line of work asks whether LLMs can assess novelty and significance rather than factual support alone. Existing evidence suggests that this remains substantially harder than constrained verification: novelty prediction benefits from retrieval and human-signal fusion, yet self-evaluation remains poorly calibrated, and AI-assisted review can distort rather than simply automate judgment~\cite{lin2025schnovel,wu2025novelty,si2025llms,liang2024monitoring,russo2025ai,ajith2026prescience}. The central limitation is that novelty and impact are not reducible to surface similarity or short-horizon scoring.

\paragraph{Progress, Limitations, and Future Directions}
Overall, LLM-as-Evaluator shows credible progress on structured verification and review assistance, but they still fall short of robust scientific judgment. The main limitations are narrow domain coverage, incomplete memorization resistance, weak calibration on novelty and methodological quality, and a lack of longitudinal evaluation in which judgments are revised as evidence accumulates. Future work should therefore emphasize cross-disciplinary coverage, stronger expert baselines, adversarial and counterfactual test settings, and protocols that track evaluative quality over time rather than at a single snapshot~\cite{song2025evaluating,wang2026frontierscience,liu2025atlas,ajith2026prescience}.

\subsection{Observations}
The literature reviewed in this section suggests that LLMs can already provide useful first-pass assistance in review generation, review aggregation, and structured claim verification. At the same time, current systems remain least reliable when evaluation requires calibrated novelty assessment, methodological scrutiny, or resistance to adversarial misrepresentation~\cite{zhou2024llm,russo2025ai,glockner2025grounding,si2025llms}. This asymmetry mirrors a broader pattern across the survey: evaluative performance improves as targets become more structured and externally grounded. At present, the Evaluator role is therefore more appropriately framed as an expert-supervised aid to validation than as an autonomous epistemic gatekeeper.

This role also occupies a distinctive position in the taxonomy because it acts as a control layer over the other three roles. If Assistant, Collaborator, and Scientist systems generate increasingly complex outputs, Evaluator systems help determine whether those outputs can be trusted, prioritized, revised, or rejected. For that reason, weaknesses in calibration and novelty judgment are disproportionately important: they do not merely affect one task, but can influence the broader human-AI research loop. The maturity of the Evaluator role may therefore condition how far the other roles can be safely and institutionally deployed.

\section{Discussion}
\label{discussion}

\subsection{Cross-Role Synthesis}
\label{sec:cross-role}

The four roles reveal that progress in AI for science is uneven: capabilities advance more rapidly in generation than in validation, and more readily in local task performance than in long-horizon reliability. Three patterns emerge that are not visible within any single role.

\paragraph{The Evaluator as a systemic constraint}
Because the Evaluator operates at the meta-level, its weaknesses propagate across the entire framework: when Evaluator systems are over-lenient, poorly calibrated on novelty, or unable to detect methodological flaws, the outputs of the other three roles cannot be filtered to scientific standards~\cite{si2025llms,russo2025ai,zhou2024llm}. The maturity of the Evaluator may therefore limit how safely the other roles can be deployed.

\paragraph{Autonomy and oversight scale asymmetrically}
As systems progress from Assistant to Scientist, autonomy increases while verifiability decreases. At the Assistant level, outputs are bounded and checkable; at the Scientist level, epistemically relevant intermediate states are distributed across time and system components~\cite{wijk2025rebench,kon2026expbench,ifargan2025autonomous}. The implication is that oversight mechanisms must become more sophisticated at the same rate as the systems they govern, not deferred as a downstream concern.

\paragraph{Benchmark evidence reflects an asymmetric human--model gap}

Table~\ref{tab:human-model-comparison} compares human and best-reported model performance on representative benchmarks spanning the four roles, each selected for its scholarly recognition, task coverage, and explicit human baseline. The pattern is asymmetric. Models surpass human baselines on both ScienceQA~\cite{lu2022learn} and BrainBench~\cite{luo2024large}, but for distinct reasons. On ScienceQA, model gains reflect broad knowledge coverage over a non-specialist baseline rather than scientific reasoning depth. On BrainBench, human participants are domain experts, yet LLMs outperform them by integrating background and method information across the full abstract, while experts tend to focus narrowly on the results section. In both cases, the format reduces scientific judgment to pattern selection that large-scale pretraining directly optimizes for.

\begin{table}[t]
\centering
\scriptsize
\caption{Human versus best reported model performance on selected benchmarks spanning the four roles. $\uparrow$: model outperforms human; $\downarrow$: model underperforms human; $\sim$: comparison is time-budget-dependent and not reducible to a single direction.}
\label{tab:human-model-comparison}
\adjustbox{max width=\textwidth}{
\begin{tabular}{lllclcc}
\toprule
\textbf{Role} & \textbf{Benchmark} & \textbf{Year} & \textbf{Metric} & \textbf{Human} & \textbf{Best Model} & \textbf{Comp.} \\
\midrule
Assistant    & ScienceQA~\cite{lu2022learn}            & 2022 & Accuracy                   & 88.40                           & 96.18 (Multimodal-T-SciQ\_Large)          & $\uparrow$   \\
Assistant    & KIWI~\cite{xu2024kiwi}                  & 2024 & Acc.\,/\,P\,/\,R\,/\,F1          & 0.80\,/\,0.63\,/\,0.90\,/\,0.75 & 0.63\,/\,0.57\,/\,0.91\,/\,0.70 (GPT-4)  & $\downarrow$ \\
Assistant    & M3SciQA~\cite{li2024m3sciqa}            & 2024 & MRR                              & 0.796                           & 0.500 (GPT-4o)                            & $\downarrow$ \\
Assistant    & PaperArena~\cite{wang2025paperarena}    & 2025 & Acc.\,/\,Steps\,/\,Eff.          & 83.50\,/\,6.52\,/\,75.48        & 38.78\,/\,8.58\,/\,45.39 (Gemini 2.5 Pro) & $\downarrow$ \\
Collaborator & PaperBench~\cite{starace2025paperbench} & 2025 & Code Dev.\,/\,Exec.\,/\,Analysis & 72.4\,/\,20.4\,/\,8.9           & 42.4\,/\,7.4\,/\,1.4 (o1, 36h)           & $\downarrow$ \\
Scientist    & BrainBench~\cite{luo2024large}          & 2024 & Accuracy                         & 63.4\%                          & 81.4\% (LLM average)                      & $\uparrow$   \\
Scientist    & RE-Bench~\cite{wijk2025rebench} & 2025 & Score@$k$ (time budget) & 8h@2 & 30min@32 (Claude), 2h@8 (o1-preview) & $\sim$ \\
Scientist    & ScienceBoard~\cite{sun2026scienceboard} & 2025 & Overall success rate             & 60.27\%                         & 15.79\% (Claude-3.7-Sonnet)               & $\downarrow$ \\
\bottomrule
\end{tabular}
}
\end{table}

The gap reverses on tasks requiring open-ended reasoning, cross-modal retrieval, or long-horizon execution. Compounding errors across multi-stage pipelines push model success rates well below those of human experts, and the deficit widens as tasks advance toward higher-order sub-tasks most closely resembling scientific judgment. A temporal asymmetry also emerges: agents make faster early progress under tight time budgets, while human experts improve substantially given more time, suggesting fundamentally different problem-solving strategies. Current LLM progress is therefore best characterized as gains on locally bounded tasks; open-ended, multi-modal, and long-horizon tasks expose a persistent shortfall, reinforcing that existing benchmarks are better suited to evaluating research assistance than autonomous scientific discovery.

\subsection{Human Agency and Institutional Implications}
\label{sec:human-agency}

\paragraph{Human contribution is reconfigured, not displaced}
The tasks that models currently perform most reliably are information-intensive, structured, and comparatively bounded: retrieval, extraction, first-pass synthesis, protocol drafting, and structured critique~\cite{luo2025llm4sr,yuan2022can}. Tasks requiring cumulative judgment, the setting of evidentiary thresholds, and accountability under uncertainty remain deeply human responsibilities. Under this interpretation, the human role is not a residual leftover after automation but the layer that determines what counts as acceptable evidence and accountable scientific progress~\cite{fugener2026roles,chen2025ai4research}.

\paragraph{The automation--augmentation distinction requires role-sensitive interpretation}
Assistant systems are the closest to bounded between-task automation, offloading discrete subtasks while humans retain interpretive authority. Collaborator systems broaden the space of candidate hypotheses and designs but do not remove the need for human filtering and validation. Scientist and Evaluator systems remain closer to within-task augmentation because the cost of undetected failure is higher and responsibility cannot be externalized to the model~\cite{zhang2025scaling,wei2025agentic}. Human agency therefore becomes more selective, not less important, as autonomy increases.

\paragraph{Institutional norms have not kept pace with capability growth}
Authorship and attribution norms are strained when a single output depends on contributions from models, retrieval systems, and coordinating agents that cannot be individually credited~\cite{lu2026ai,ifargan2025autonomous}. Peer review presupposes that evaluative judgments can be traced to specific claims and contested; LLM-generated reviews still lack that claim-level grounding~\cite{yuan2022can,zhu-etal-2025-deepreview}. Sustained reliance on Assistant-style systems may reduce routine cognitive labor while raising the stakes of higher-order domain judgment~\cite{messeri2024artificial}. Disclosure standards, review transparency requirements, and norms governing higher-autonomy systems remain underdeveloped relative to the pace of capability growth.

\subsection{Challenges}
\label{sec:challenges}

\subsubsection{System-Level Failures Beyond Component Capability}
\label{sec:challenge-system}

\paragraph{Local competence does not ensure global reliability}
Scientific AI systems are increasingly implemented as compositions of interacting modules, yet failures arise less from the absence of capability in any single module than from breakdowns at their interfaces. In peer review simulation, agent interaction can reproduce or amplify collective bias even when individual reviewer agents appear locally calibrated~\cite{Jin2024AgentReviewEP,goyal2026scholarpeer}. In autonomous research pipelines, misalignment between planning and execution produces errors that are difficult to localize to any single step~\cite{boiko2023autonomous,mandal2025evaluating}. Whether inter-agent communication relies on natural language or structured state representations has direct consequences for the diagnosability of such failures.

\paragraph{Current memory mechanisms cannot support long-horizon task execution}
Scientific innovation unfolds iteratively: hypotheses are revised across rounds, experimental strategies depend on prior failures, and evaluative judgments shift as new evidence accumulates. Current architectures address this through finite context windows or retrieval-based memory, neither of which provides the structured, updateable state that such workflows demand. Without persistent memory, systems may revisit exhausted regions of the hypothesis space rather than build on previous attempts~\cite{wijk2025rebench,kon2026expbench}, and fail to track how judgments evolve across review rounds~\cite{zhang2025re}. The core architectural challenge is therefore not output generation, but state preservation, component coordination, and error recovery across extended task execution.

\subsubsection{Generalization and Evaluation}
\label{sec:challenge-eval}

\paragraph{Scientific usefulness requires more than broad fluency}
LLMs often display wide conceptual coverage across disciplines, but scientific usefulness depends on internalizing field-specific standards for novelty, feasibility, rigor, and evidentiary sufficiency. This gap manifests across all four roles: Assistant systems degrade in low-resource or highly specialized domains~\cite{wei2024pubtator,duan2025scinlp}; Collaborator systems produce hypotheses plausible in form but inconsistent with disciplinary norms~\cite{liu2025hypobench,yang2025moosechem}; Scientist systems generate plans that violate domain-specific safety or reproducibility constraints~\cite{mandal2025evaluating,darvish2025organa}; Evaluator systems apply review criteria that do not transfer reliably across fields~\cite{yan2025surveyforge,si2025llms}. Whether apparent generality reflects genuine scientific transfer or superficial pattern familiarity remains unresolved.

\paragraph{Current metrics are poorly aligned with scientific value}
Many benchmarks overlap with pretraining data, so high performance may reflect memorization rather than genuine reasoning or discovery capability~\cite{jansen2024discoveryworld,zheng2026newtonbench}. Similarity-based metrics do not capture novelty, feasibility, or downstream viability in hypothesis generation~\cite{guo2025ideabench,wang2024scimon}; single-round annotation agreement does not capture how well Evaluator systems track evolving methodological standards~\cite{russo2025ai,liang2024monitoring}; and workflow completion rates do not indicate the epistemic quality of resulting claims. What is missing is a class of evaluation protocols that jointly assess novelty, grounding, reproducibility, and cross-domain transfer under conditions resembling actual scientific practice.

\subsubsection{Opacity and Homogenization as Systemic Risks}
\label{sec:challenge-accountability}

\paragraph{Agentic opacity creates irreducible provenance gaps}
Under agentic conditions, a final result may depend on stochastic generation, external retrieval, tool calls, and inter-agent communication, yet the intermediate states governing the output are rarely preserved in inspectable form~\cite{lu2026ai,ifargan2025autonomous}. This opacity makes it difficult to verify which sources support which claims or which component introduced a given error, allowing fabricated citations, unsupported synthesis, and unverifiable hypotheses to propagate through downstream interpretation and citation~\cite{luo2025llm4sr,zhang2024comprehensive}.

\paragraph{Shared training priors risk epistemic homogenization}
LLMs across all four roles share overlapping training corpora, encoding convergent priors about what constitutes a plausible hypothesis, a rigorous methodology, or a publishable contribution. Large-scale empirical analysis demonstrates that AI-assisted research narrows topical diversity and increases clustering around already-prominent questions, even as it amplifies individual researcher productivity~\cite{hao2026artificial}. When generation and evaluation are both governed by models trained on the same literature, unconventional work is systematically discounted in favor of consensus-aligned output~\cite{messeri2024artificial}. Epistemic homogenization is therefore a population-level effect: independent review processes develop correlated blind spots, and the hypothesis space explored by the field contracts because the same distributional prior governs both ideation and assessment.

\subsection{Future Directions}
\label{sec:future}

\subsubsection{Reliable and Inspectable System Design}

Inspectability and provenance should be first-order design requirements.
Intermediate representations, including hypothesis structures, evidence states, sub-plans, execution traces, and agent communications, should be preserved alongside final outputs so that errors in extended workflows can be localized and results reconstructed by human overseers~\cite{shao2025sciscigpt,chen2025ai4research}. Beyond runtime auditability, systems should produce structured records of retrieved sources, model versions, tool calls, and inter-agent exchanges that materially shaped the final output~\cite{lu2026ai,ifargan2025autonomous}. Such records are necessary not only for governance but for reproducibility: without them, retraction, correction, and priority adjudication lose their evidentiary foundation.

\subsubsection{Trajectory-Based Evaluation Protocols}

Evaluation should reflect the temporal structure of scientific work.
Point-in-time measurement should give way to trajectory-based protocols: for Collaborator systems, hypothesis generation should be connected to downstream experimental testing and revision within the same evaluation pipeline; for Scientist systems, evaluation should measure error recovery and strategic adaptation across multi-step workflows rather than single-pass completion; for Evaluator systems, judgments should be calibrated against longitudinal expert assessments that track opinion change across revision rounds, not single-round annotation agreement~\cite{ajith2026prescience,song2025evaluating,russo2025ai,liang2024monitoring,he2026trajectbencha}.

\subsubsection{Domain-Specific Scientific Foundation Models}

Domain-specific competence requires scientific foundation models.
Current evaluations concentrate on computer science and biomedicine; expanding coverage requires benchmarks that encode each field's criteria for novelty, feasibility, and evidentiary sufficiency~\cite{jansen2024discoveryworld,zheng2026newtonbench}. The deeper challenge is epistemic: LLMs acquire broad instrumental knowledge through next-token prediction, but domain-specific scientific reasoning requires internalized world models of how interventions, mechanisms, and constraints operate within a field~\cite{yildirim2024task}. Scientific foundation models trained on domain corpora that include experimental records, simulation trajectories, and failure data offer a more principled path toward this competence; progress in chemistry and materials science already demonstrates that domain-specific pretraining yields models whose outputs respect disciplinary constraints in ways that general-purpose LLMs do not~\cite{liu2025foundation}. For Collaborator and Scientist systems, this gap is consequential: a hypothesis that is linguistically well-formed but violates tacit experimental norms will not be flagged by domain-agnostic metrics~\cite{liu2025hypobench,darvish2025organa}.

\subsubsection{Governance and Epistemic Diversity}

Governance and diversity mechanisms should be embedded in workflow design.
Disclosure norms should distinguish among different forms of AI participation across the four roles rather than treating all AI use as equivalent~\cite{chen2025ai4research,zhang2025scaling}. To counteract epistemic homogenization, pipeline design should deliberately introduce model diversity at the evaluation stage, drawing on heterogeneous model families or structured adversarial review, so that correlated blind spots induced by shared training priors do not uniformly suppress unconventional contributions~\cite{messeri2024artificial}.

\section{Conclusion}

This survey examined the role of large language models in scientific innovation through a four-role framework comprising \textbf{Assistant}, \textbf{Collaborator}, \textbf{Scientist}, and \textbf{Evaluator}. Our main argument is that these roles cannot be understood by autonomy alone, because each combines different cognitive demands, forms of scientific innovation, failure modes, and oversight requirements. In this sense, the most important near-term question is not whether LLMs will become ``scientists'' in a singular sense, but how different configurations of models, tools, and supervision should be matched to different scientific functions. We therefore view progress in AI for science not simply as a matter of stronger model capability, but as a matter of aligning system design, evaluation, and institutional safeguards with the epistemic standards that make scientific practice reliable.

\printbibliography
\end{document}